\documentclass[11pt]
{revtex4}

\usepackage{amscd}
\usepackage{amsmath}

\begin{document}

\title{Parabolic Coordinates  and the Hydrogen Atom in Spaces $H_{3}$ and  $ S_{3}$
}
\author{V.M. Red'kov }

\author{E.M. Ovsiyuk}

\email{ redkov@dragon.bas-net.by, e.ovsiyuk@mail.ru}

\affiliation{Institute of Physics, NAS of Belarus \\
Mozyr State Pedagogical University  }

\begin{abstract}

The Coulomb  problem for Schr\"{o}dinger equation is examined, in
spaces of constant curvature,
 Lobachevsky $H_{3}$ and Riemann $S_{3}$ models, on the base of generalized parabolic coordinates.
 In contrast to the hyperbolic case, in spherical space $S_{3}$ such
  parabolic coordinates turn to be complex-valued, with additional constraint on them.
The technique of the use of such real and complex coordinates in
two space models within the method of separation of variables  in
Schr\"{o}dinger equation with Kepler potential is developed in
detail; the energy spectra and corresponding wave functions  for
bound states have been constructed in explicit form for both
spaces; connections with Runge-Lenz operators in both curved space
models are described.
\end{abstract}


\maketitle

\newcommand{\N}{N\raise.7ex\hbox{\underline{$\circ $}}$\;$}

\section{ Introduction}

Quantum mechanics had been started with the theory of the
hydrogen atom, so  considering  the Quantum mechanics in
Riemannian spaces it is first natural step to turn to just this
system. A common  quantum-mechanical hydrogen atom model is based
materially on  the assumption of the Euclidean character of the
physical 3-space geometry. In this context,  natural questions
arise:  what in  such a model description is determined by this
assumption, and which changes  will be entailed by allowing for
other spatial geometries: for instance. Lobachevsky's  $H_{3}$,
Riemann's $S_{3}$, or de Sitter geometry. The question is of
fundamental significance, even beyond  its possible  experimental
testing.

Firstly, the  hydrogen atom in 3-dimensional  space of constant
positive curvature $S_{3}$  was considered by Schr\"{o}dinger
\cite{1940-Schrodinger}. He had been studied the so-called
factorization method in quantum mechanics; in particular,
application of this techniques to discrete  part of the energy
spectrum for hydrogen atom had been elaborated. An idea was to
modify the basic  atom system so as to cover all the energy
spectrum including the region $E>0$ as well. However, mere placing
of the atom system inside a finite box  in order  to make  the
whole energy spectrum discrete did not seem attractive, so
Schr\"{o}dinger had placed the atom into the curved background of
the Riemann space model  $S_{3}$. Due its compactness, the
spherical Riemann model may simulate the effect of the finite box
 -- see Schr\"{o}dinger \cite{1940-Schrodinger} and  Stevenson \cite{1941-Stevenson}.

In spherical coordinates of $S_{3}$
\begin{eqnarray}
dl^{2} = d \chi^{2} + \sin^{2}\chi (d\theta^{2} +
\sin^{2}d\phi^{2}) \;
\nonumber
\end{eqnarray}
\noindent the Schr\"{o}dinger Hamiltonian in dimensionless units
has the form
\begin{eqnarray}
H = -{1 \over 2} \; {1 \over \sqrt{g}} \; {\partial \over \partial
x^{\alpha}} \; \sqrt{g} \; g^{\alpha \beta}\; {\partial \over
\partial x^{\alpha}} \; -\; {e \over \mbox{tan}\;  \chi }  \;  ;
\nonumber
\end{eqnarray}
\noindent $\rho$ is a curvature radius, a unite for length; $M$ is
a mass of the electron; $\hbar^{2} /M\rho^{2}$ is taken as a unit
for energy;  $e = {\alpha \over \rho } /  { \hbar^{2} \over M
\rho^{2} } $ stands for a Coulomb interaction constant; the sign
at  $e / \mbox{tan}\; \chi $  corresponds to the attracting
Coulomb force. The energy spectrum is entire discrete and given by
\begin{eqnarray}
\epsilon_{n} = - {e^{2} \over 2 n^{2} } + {1 \over 2} \; (n^{2}
-1) \; ,
 \;\;  n = 1,2, 3, ...
\nonumber
\end{eqnarray}

Hydrogen atom in the Lobachevsky space  $H_{3}$  was considered
firstly by
 Infeld and Shild  \cite{1945-Infeld-Schild}
\begin{eqnarray}
dl^{2} = d \chi^{2} + \mbox{cosh}^{2}\chi (d\theta^{2} +
\sin^{2}d\phi^{2}) \;  , \nonumber
\\
H = -{1 \over 2}  {1 \over \sqrt{g}}  {\partial \over \partial
x^{\alpha}} \; \sqrt{g} \; g^{\alpha \beta}  {\partial \over
\partial x^{\alpha}}  -  {e \over \mbox{tanh}\;  \chi }  \; .
\nonumber
\end{eqnarray}
\noindent Energy spectrum contains a discrete and continuous
parts. The number of discrete levels is finite,  they are
specified  by
\begin{eqnarray}
- \; {e^{2} \over 2} \leq \epsilon \leq ({1\over 2}  - e ) \; ,
\qquad \epsilon_{n} = - {e^{2} \over 2 n^{2} } - {1 \over 2} \;
(n^{2} -1) \; ,\;\;
    n = 1,2, 3, ...,N \; .
\nonumber
\end{eqnarray}
\noindent   In the region  $\epsilon \geq ({1\over2} - e)$ the
energy spectrum is continuous.

Thus, the models of the hydrogen atom in Euclid, Riemann, and
Lobachevsky spaces significantly differ from each other,
 which is the result of differences in three spatial  geometries:
  $E_{3},\; H_{3},\; S_{3}$. To present time, we see a plenty of investigations
  on this matter:

Higgs \cite{1979-Higgs}, Leemon \cite{1979-Leemon}, Kurochkin --
Otchik \cite{1979-Kurochkin-Otchik}, Bogush -- Kurochkin -- Otchik
\cite{1980-Bogush-Kurochkin-Otchik}, Parker \cite{1980-Parker},
\cite{1981-Parker}, Ringwood -- Devreese
\cite{1980-Ringwood-Devreese}, Kobayshi \cite{1980-Kozo}, Bessis
-- Bessis -- Shamseddine \cite{1982-Bessis-Bessis-Shamseddine},
Grinberg -- Maranon -- Vucetich
\cite{1983-Grinberg-Maranon-Vucetich}, Bogush -- Otchik -- Red'kov
\cite{1983-Bogush-Otchik-Red'kov}, Bessis -- Bessis -- Shamseddine
\cite{1984-Bessis-Bessis-Shamseddine(1)},\cite{1984-Bessis-Bessis-Roux(2)},
\cite{1984-Bessis-Bessis(3)}, Chondming -- Dianyan
\cite{1984-Chondming-Dianyan}, Xu -- Xu \cite{1984-Xu-Xu},
Melnikov -- Shikin \cite{1985-Melnikov-Shikin}, Shamseddine
\cite{1986-Shamseddine}, Otchik -- Red'kov
\cite{1986-Otchik-Red'kov}, Barut -- Inomata -- Junker
\cite{1987-Barut-Inomata-Junker}, Bessis -- Bessis -- Roux
\cite{1988-Bessis-Bessis-Roux}, Bogush -- Otchik -- Red'kov
\cite{1988-Bogush-Otchik-Red'kov.}, Gorbatsievich -- Priebe
\cite{1989-Gorbatsievich-Priebe}, Groshe \cite{1990-Groshe}, Barut
-- Inomata -- Junker \cite{1990-Barut-Inomata-Junker}, Katayama
\cite{1990-Katayama}, Chernikov \cite{1992-Chernikov}, Mardoyan --
Sisakyan \cite{1992-Mardoyan-Sisakyan}, Granovskii -- Zhedanov --
Lutsenko \cite{1992-Granovskii-Zhedanov-Lutsenko}, Kozlov -- Harin
\cite{1992-Kozlov-Harin}, Vinitskii -- Marfoyan -- Pogosyan --
Sisakyan -- Strizh
\cite{1993-Vinitskii-Marfoyan-Pogosyan-Sisakyan-Strizh},
Shamseddine \cite{1997-Shamseddine}, Bogush -- Kurochkin -- Otchik
\cite{1998-Bogush-Kurochkin-Otchik}, Otchik \cite{1999-Otchik},
Nersessian -- Pogosyan \cite{2001-Nersessian-Pogosyan}, Red'kov
\cite{2001-Red'kov}, Bogush -- Kurochkin -- Otchik
\cite{2003-Bogush-Kurochkin-Otchik}, Kurochkin -- Otchik --
Shoukavy \cite{2006-Kurochkin-Otchik-Shoukavy}, Kurochkin --
Shoukavy \cite{2006-Yu. Kurochkin-Shoukavy}, Bogush -- Otchik --
Red'kov \cite{2006-Bogush-Otchik-Red'kov}, Bessis -- Bessis
\cite{1979-Bessis-Bessis},  Iwai \cite{1982-Iwai}, Cohen -- Powers
\cite{1982-Cohen-Powers}, Ovsiyuk \cite{1982-Ovsiyuk}.

\section{Coordinates in space $H_{3}$  and $S_{3}$}

In Euclidean 3-dimension space $E_{3}$ there exist 11  coordinate
systems
\cite{Eisenhart},
 allowing for the complete  separation of variables in the Helmholtz  equation
\begin{eqnarray}
 \left ( {1 \over \sqrt{g}}  {\partial \over \partial x^{\alpha}}
 \sqrt{g}  g^{\alpha \beta}\; {\partial \over
\partial x^{\alpha}} + \lambda \right  )  \Phi (x^{1}, x^{2}, x^{3}) = 0 \; ;
\nonumber
\label{1.1}
\end{eqnarray}
\noindent $g^{\alpha \beta}(x)$ stands for the metric tensor of
space $E_{3}$ specified for  curvilinear coordinates
$(x^{1},x^{2},x^{3})$. Solution of the same problem for spaces of
constant positive and negative curvature, Riemannian  $S_{3}$ and
Lobachevsky $H_{3}$
 models  was given by Olevsky in \cite{Olevsky}. It was established that  there exist 34
 such  coordinate systems for hyperbolic space  $H_{3}$, whereas in the case of spherical model
 $S_{3}$  the number  of those systems is only
  6. Extension of the analysis  to the complex sphere was given by
Kalnins and Miller in \cite{1978-Kalnins-Miller}. Also see
\cite{Herranz-Ballesteros-2006, 1995-Grosche-Pogosyan-Sissakian,
1997-Grosche-Pogosyan-Sissakian,1997-Grosche-Pogosyan-Sissakian,
2009-Pogosyan-Yakhno}.


The above asymmetry between  $H_{3}$ and $S_{3}$  may be seen   as
even more   strange  if one  calls the known relations   of these
models  through  the analytical continuation
\begin{equation}
\begin{split}
H_{3} \qquad  x_{0}^{2} -  x_{1}^{2}- x_{2}^{2} - x_{3}^{2}  = + 1 \;
\; ,
\\
S_{3} \qquad  y_{0}^{2} +  y_{1}^{2}+ y_{2}^{2} + y_{3}^{2}  = + 1
\; ;
\end{split}
\label{1.2}
\end{equation}
\noindent   the curvature radius  $R$ is taken as a unit for
length.

The  asymmetry  of the models $H_{3}$ and $S_{3}$  with respect to
coordinate systems finds its logical  corollary  when turning to
the study of the quantum mechanical  model for a hydrogen atom  on
the background of  a curved space.  In particular,
an additional degeneracy like in the  case of flat space was
observed,
 which presumes  existence of  a hidden symmetry in the (curved space)  problem.
 In  \cite{1979-Higgs, 1979-Leemon, 1979-Kurochkin-Otchik, 1980-Bogush-Kurochkin-Otchik},
  the symmetry operators accounting for such additional degeneracy in Kepler
  problem on curved space background ware found for both model $H_{3}$ and $S_{3}$,
 and    an analog
  of the conventional Runge -- Lenz vector in the flat space  was constructed.

  Connection between  the Runge -- Lenz operator $\vec{A}$  in the quantum Kepler problem
and parabolic coordinates in   Euclidean space is well known: by
 solving the   Schr\"{o}dinger equation in these coordinates  the  eigenfunctions of
 the third component  $A_{3}$
   arise [13]. Analogous situation exists in the hyperbolic
    space $H_{3}$  but not in  in the spherical $S_{3}$. In the Lobachewsky space,
    among 34 coordinates established by Olevsky \cite{Olevsky}
    one may select  one  special case of
 parabolic     system of coordinates in $H_{3}$, in which
  the Schr\"{o}dinger equation  allows the separation of variables and the  wave
 functions arisen turn out  to be  eigenfunctions of the operator $B = A_{3} +L^{2}$.
 Among six coordinate systems mentioned  in [4] an analog of parabolic coordinates
  is  not encountered.

If one looks at  34 and 6  systems in $H_{3}$ and $S_{3}$
respectively, one can note that all six ones from $S_{3}$ have
their counterparts in $H_{3}$.
    The  main  purpose   of  the present paper consists is the search  of some counterparts
    of remaining $34-6=28$  systems. It turns  out that such 28  systems in $S_{3}$
     can be constructed, but they  should be complex-valued; to preserve real nature of the geometrical space  one must impose
    additional  restrictions  including complex conjugation.

In particular, the complex  analog for parabolic coordinates in
space  of the positive curvature $S_{3}$  can be introduced and
used in studying the quantum mechanical Kepler problem in this
space (this possibility was partly studied  before in \cite{1983-Bogush-Otchik-Red'kov}, \cite{1986-Otchik-Red'kov},
    \cite{2006-Bogush-Otchik-Red'kov}).

Let us start with the following fact: from the the metrics in
Lobachevsky space (note $\chi \in [0, + \infty ) )$
\begin{eqnarray}
dl^{2} =  d \chi ^{2} + \sinh ^{2} \chi \; (d \theta ^{2} + \sin
^{2} \theta \; d\phi ^{2})     \label{1.3}
\end{eqnarray}
\noindent   by means of the change
 $ \chi  \rightarrow  i \chi \; ,  \; \sinh \chi  \rightarrow
 i \sin \chi $ one can  obtain the corresponding metrics of
the Riemannian space (note that $\chi \in [0, \pi ]$)
\begin{eqnarray}
 dl^{2} = -  [ \; d
\chi ^{2} + \sin ^{2} \chi \; (d \theta ^{2} + \sin ^{2} \theta \;
d\phi ^{2})\;] \;  .
\label{1.4}
\end{eqnarray}
\noindent This  simple observation on $H_{3}-S_{3}$ connection
leads us to interesting    consequences. Indeed,  let us compare,
for instance,   wave functions and spectra   for hydrogen atom in
spaces of negative and constant curvature
\begin{eqnarray}
\underline{H_{3}},  \qquad
 \Psi _{nlm} (\chi, \theta, \phi ) = N
S(\chi)  Y_{lm}(\theta , \phi )  \; ,
\nonumber
\\
S(\chi) = \sinh ^{l} \chi  \exp \left[\left(n-l-1 - {e \over n}\right) \chi \right]
\nonumber
\\
\times
F\left({e \over n} + l + 1 , l - n + 1, 2l+2; 1 - e^{-2\chi}\right) \; ,
\nonumber
\label{1.5}
\\
\epsilon_{n} = - {e^{2} \over 2 n^{2} } - {1 \over 2}  (n^{2}
-1) \; ;
\label{1.6}
\end{eqnarray}
\begin{eqnarray}
\underline{S_{3} }, \qquad  \Psi _{nlm} (\chi, \theta, \phi
) = K  S(\chi)  Y_{lm}(\theta , \phi )  \; ,
\nonumber
\\
S(\chi) = \sin ^{l} \chi  \exp \left[\left(i (n-l-1) - {e \over n}\right) \chi \right]
\nonumber
\\
 \times F\left (-i {e \over n} + l + 1 , l - n + 1, 2l+2; 1 - e^{-2i\chi}\right)
\; ,
\nonumber
\\
\epsilon_{n} = - {e^{2} \over 2 n^{2} } + {1 \over 2}  (n^{2} -1)
\; , \; e = {\alpha \over  R } /  { M \hbar^{2} \over R^{2} } \; ;
\label{1.8}
\end{eqnarray}

\noindent quantity ($M\hbar^{2} /R^{2})$ provides us with natural
unit for energy, $e$ is a dimensionless parameter  characterizing
intensity of the Coulomb interaction. One may readily note that
these two solutions turn into each other at the following formal
replacement
\begin{eqnarray}
\chi \; \longrightarrow \; i \; \chi \; , \;\; e \longrightarrow
-i\; e \; , \;\; \epsilon \longrightarrow - \; \epsilon \; .
\label{1.9}
\end{eqnarray}

This example indicates that  the relation  between $H_{3}$ and
$S_{3}$ reflected by substitution   $\chi \rightarrow i \chi$ is
meaningful. In the context  of the described  above situation with
coordinate systems in $H_{3}$ and $S_{3}$,  let us  make use of
this correspondence ($ \chi \; \longrightarrow \; i \; \chi$) as
follows.

 Let in the Lobachevsky space  $H_{3}$ be chosen  a coordinate
system $(\rho_{1},\; \rho_{2}, \; \rho_{3} )$ (one of those 34
found by Olevsky), then as a first step one has to  establish
connection of  such a system with spherical one:
\begin{eqnarray}
\rho _{k} = f_{k}( \chi, \; \theta, \; \phi ) \; ,
\label{1.10}
\end{eqnarray}

\noindent and  a second step  is to  introduce a corresponding
coordinate system in the space $S_{3}$ through the formal change
$\chi \longrightarrow i\;\chi$:
\begin{eqnarray}
\rho _{k} = f_{k}( i \chi, \; \theta, \; \phi ) \; .
\label{1.11}
\end{eqnarray}

With  help of this prescription one can determine 34 coordinate
systems  in space $S_{3}$ in comparison  to six ones given in \cite{Olevsky}. It turns out that  28 new (added) coordinate systems are
complex-valued and  therefore  additional restrictions  should be
imposed
 which involve  complex conjugation.
All these  extra  coordinate systems  permit the full separation
of variables in the Helmholtz  equation on the sphere $S_{3}$.

Below, only one example of such coordinates, analog of   the
parabolic ones  in space $H_{3}$, will be examined in detail  and
applied to the study of the quantum-mechanical Kepler problem on
the sphere $S_{3}$.

\section{
Parabolic coordinates in space models  $S_{3}$, $H_{3}$ }

In  \cite{Olevsky}, the following coordinate system (the
case  $XXV$) in Lobachevsky space had been given
\begin{widetext}
\begin{eqnarray}
dl^{2} =  \; {(\rho_{1} - \rho_{2}) \over 4 (\rho _{1} -
a)(\rho_{1} - b)^{2}} \; d \rho^{2}_{1} \; + \; {(\rho_{2} -
\rho_{1}) \over 4 (\rho _{2} - a)(\rho_{2} - b)^{2}} \; d
\rho^{2}_{2} \; - \; (\rho_{1} - a)(\rho_{2} - a) d \rho_{3}^{2}
\; ,\label{2.1a}
\end{eqnarray}
\end{widetext}
\noindent where  $(\rho_{1},\; \rho_{2}, \; \rho_{3} )$ are
connected with the four quasi-Cartesian  coordinates  $(x_{0},
x_{1},x_{2}, x_{3})$
\begin{eqnarray}
-(x_{0})^{2} + (x_{1})^{2} + (x_{2})^{2} + (x_{3}) ^{2} = -1 \;
,\qquad    x_{0}  > +1
\nonumber
\end{eqnarray}
\noindent by the formulas
\begin{equation}
\begin{split}
{x_{2} \over x_{1}} = \mbox{tan}\;  [(a-b) \rho_{3}] \; , \quad b
< \rho_{1} < a < \rho_{2}  ,
\\
{x_{1}^{2} +  x_{2}^{2}  \over \rho_{i} - a } + {x_{3}^{2} -
x_{0}^{2}  \over \rho_{i} - b } + {(x_{3} - x_{0})^{2} \over (\rho
_{i}  - b )^{2}}
 = 0 \;\;\;  ( i = 1,2 ) \; .
\end{split}
\label{2.1b}
\end{equation}
\noindent With the notation
\begin{eqnarray}
 x_{1}^{2} +  x_{2}^{2}  = \sigma^{2}
\; , \;  x_{3} - x_{0} = U \; , \;  x_{3} + x_{0} = V \; ,
\nonumber
\end{eqnarray}
\noindent and  $a = +1, \; b = 0 $, eq.  (\ref{2.1b}) gives
\begin{eqnarray}
\sigma^{2} + UV = -1 \; , \qquad
  x_{1} = \sigma \cos \rho_{3} \;
,\qquad  x_{2} = \sigma \sin \rho_{3} \; ,
\nonumber
\\
{ \sigma^{2} \over \rho_{1} -1 } + {UV \over \rho_{1}} + {U^{2}
\over \rho_{1}^{2}} = 0 \; , \qquad
 { \sigma^{2} \over \rho_{2}
-1 } + {UV \over \rho_{2}} + {U^{2} \over \rho_{2}^{2}} = 0 \; .
\nonumber
\end{eqnarray}
\noindent  Combining two last equations results in
\begin{eqnarray}
( {\rho_{1} \over \rho_{1} -1 }  -  {\rho_{2} \over \rho_{2} -1 }
)\; \sigma^{2}  + ( {1 \over \rho_{1} }  -  {1 \over \rho_{2} }
)\;
 U^{2} = 0\; ,
\nonumber
\\
( {\rho_{1}^{2} \over \rho_{1} -1 }  - {\rho_{2}^{2} \over
\rho_{2} -1 } )\; \sigma^{2}  + ( \rho_{1}  - \rho_{2} ) \; U V =
0  \; .
\nonumber
\end{eqnarray}
\noindent  From this, taking into account   $\sigma^{2} + UV =
-1$, we arrive at
\begin{eqnarray}
{U \over V} = { \rho_{1} \rho_{2} \over \rho_{1} \rho_{2} -
\rho_{1} -  \rho_{2}  } \; , \; \; UV = \rho_{1} \rho_{2} -
\rho_{1} -  \rho_{2}  \; ,
\nonumber
\end{eqnarray}
\noindent and further
\begin{eqnarray}
U^{2} = \rho_{1} \rho_{2} \; , \;\;
 V = U \; {
\rho_{1} \rho_{2} - \rho_{1} -  \rho_{2}  \over \rho_{1} \rho_{2}
} \; .
\nonumber
\end{eqnarray}
\noindent Thus, for  $U, V,  \sigma $ we have found (the
Lobachevsky space is realized on the branch $x_{0} > +1$, so that
$(x_{3} - x_{0}) \leq  0\; $)
\begin{eqnarray}
U = x_{3} - x_{0} = - \sqrt{\rho_{1} \rho_{2}} \;\; ,
\qquad
V =
x_{3} + x_{0} = { \rho_{1} +  \rho_{2} - \rho_{1} \rho_{2} \over
\sqrt{ \rho_{1} \rho_{2} } }   \; ,
\nonumber
\\
\sigma = \sqrt{-1 - UV }= \sqrt{-(1-\rho_{1})(1- \rho_{2})} \; .
\nonumber
\end{eqnarray}
\noindent Explicit formulas relating $\rho_{1},\rho_{2},\rho_{3}$
with  Cartesian coordinates $(x_{0},x_{l})$ look as
\begin{eqnarray}
\begin{split}
x_{1} = \sqrt{-(1-\rho_{1})(1-\rho_{2})}\; \cos  \rho_{3} \;\; ,
\\
x_{2} = \sqrt{-(1-\rho_{1})(1-\rho_{2})}\; \sin  \rho_{3} \;\; ,
\\
x_{3} = {\rho_{1} + \rho_{2} - 2 \rho_{1} \rho_{2} \over 2
\sqrt{\rho_{1} \rho_{2} }} \;\;  , \;    x_{0} = {\rho_{1} +
\rho_{2} \over  2  \sqrt{ \rho_{1} \rho_{2} } }\; ;
\end{split}
\label{2.2}
\end{eqnarray}
\noindent  and the  inverse formulas are
\begin{eqnarray}
\begin{split}
\rho_{1} = {x_{0} - x_{3}  \over  x_{0} +  x } \; , \;
\rho_{2} = {x_{0} - x_{3}  \over  x_{0} -  x } \;\; ,
\\
\rho_{3} = \mbox{arctan} { x_{2}  \over  x_{1} } \;      , \;
x = \sqrt{x_{1}^{2} + x_{2}^{2} + x_{3}^{2}} .
\end{split}
\label{2.3}
\end{eqnarray}

Now, instead of the introduced $\rho_{1},\rho_{2},\rho_{3}$ one
can   define other coordinates which  behave    simply
 in the limit  $R \rightarrow \infty $  (the curvature vanishes). Such a limiting
  procedure    for spherical  coordinates of  the hyperbolic space $H_{3}$ with metrics
\begin{eqnarray}
dl^{2} = \rho ^{2} \; [ \; d \chi ^{2} + \sinh ^{2} \chi \; (d
\theta ^{2} + \sin ^{2} \theta \; d\phi ^{2})\;]
\nonumber
\end{eqnarray}
\noindent  going over into spherical ones of the flat space
$E_{3}$
\begin{eqnarray}
\lim_{\rho \rightarrow \infty} (\rho \chi ) = r \; , \;
\lim_{\rho \rightarrow \infty} (\rho \sinh \chi ) = r \; .
\label{2.4a}
\end{eqnarray}
\noindent Eliminating   $x_{0}$ as follows
\begin{eqnarray}
q_{l} = {x_{l} \over x_{0}} = {x_{l} \over + \sqrt{1 + x^{2}}} \;
, \; q_{l} =  \mbox{tanh}\;  \chi \; n_{l} \, , \qquad
 n_{l} =
(\sin \theta \cos \phi, \sin \theta \sin \phi, \cos \theta )
\nonumber
\end{eqnarray}
\noindent we note that  when $R \rightarrow \infty$ the
coordinates $q_{l}$  reduce to
\begin{eqnarray}
\lim_{\rho \rightarrow \infty } (\rho q_{l}) = \lim_{\rho
\rightarrow \infty } (\rho \; \mbox{tanh}\; \chi \; n_{l}) = r \;
n_{l} \; .
\label{2.4b}
\end{eqnarray}
\noindent So, to have coordinates  with   known and understandable
behavior in the limit $R \rightarrow \infty $ we  define   new
coordinates $t_{1},t_{2},\phi$
\begin{eqnarray}
\begin{split}
t_{1} = 1 - \rho_{1} = {q_{3} + q \over 1 + q }\; , \;
 t_{2} =
1 - \rho_{2} = {q_{3} - q \over 1 - q } \;  , \;
 \phi =
\rho_{3} = \mbox{arctan} \; {q_{2} \over q_{1}} \; ;
\end{split}
\label{2.5a}
\end{eqnarray}
\noindent in the limit of the flat space they provide us with  the
known parabolic  coordinates $(\xi, \eta, \phi)$
\begin{equation}
\begin{split}
\lim_{\rho \rightarrow \infty } (\rho t_{1}) = z + r = \xi \; ,
\qquad
 \lim_{\rho \rightarrow \infty } (\rho t_{2}) = z - r = -
\eta \;\; .
\end{split}
\label{2.5b}
\end{equation}

The metrics  (\ref{2.1a})  in coordinates   $(t_{1},t_{2},\phi)$
takes the form
\begin{eqnarray}
dl^{2} =  {t_{1} - t_{2} \over 4 t_{1} (1 - t_{1})^{2}} \;
dt_{1}^{2} \;+\;  {t_{2} - t_{1} \over 4 t_{2} (1 - t_{2})^{2}} \;
dt_{2}^{2}  \;
- t_{1} t_{2} \; d \phi^{2} \;  ,
\nonumber
\\
0 \leq t_{1} \leq 1 \, \, , \qquad
\;  t_{2} \leq 0 \;\; , \qquad  0 \leq \phi \leq 2\pi \; . \label{2.6}
\end{eqnarray}

Now, with the help of the rules  (\ref{1.10}) and (\ref{1.11}), one can define
 corresponding parabolic coordinates $t_{1}, t_{2}$ on the sphere $S_{3}$.
To this end,  coordinates  $(t_{1}, t_{2})$ in  $H_{3}$ must be
expressed in terms of spherical ones $(\chi, \theta)$
\begin{equation}
\begin{split}
t_{1} = (1 + \cos \theta )\; { \mbox{tanh}\;  \chi \over 1 +
\mbox{tanh}\; \chi } \; , \qquad
 t_{2} = (1 - \cos \theta )\; { -
\mbox{tanh}\;  \chi \over 1 - \mbox{tanh}\;  \chi } \;
\end{split}
\label{2.7}
\end{equation}
\noindent from whence we get defining relations  for corresponding
coordinates  in $S_{3}$
\begin{equation}
\begin{split}
t_{1} = (1 + \cos \theta )\; { i \; \mbox{tan}\; \chi \over 1 +
i\; \mbox{tan}\;  \chi } \; , \qquad
 t_{2} = (1 - \cos \theta )\;
{ - i \;\mbox{tan}\; \chi \over 1 - i \mbox{tan}\;  \chi } \; .
\end{split}
\label{2.8a}
\end{equation}
\noindent Take special notice that  $(t_{1}$ and $t_{2})$  in
(\ref{2.8a})  are complex-valued expressed  through  two real   $(\chi,
\theta)$. The inverse formulas are readily found
\begin{eqnarray}
1 + \cos \theta = t_{1} (1 + {1 \over i q})  \;, \qquad  1 - \cos
\theta = t_{1} (1 - {1 \over i  q}) \; ,
\nonumber
\\
\cos \theta  = {t_{1} + t_{2} - 2 t_{1} t_{2} \over    t_{1} -
t_{2} }\;  , \qquad  iq = {t_{1} - t_{2} \over 2 - t_{1} - t_{2} } \; .
\label{2.8b}
\end{eqnarray}

So defined  parametrization of  $S_{3}$ by  coordinates $t_{1},
t_{2}$ can be additionally  detailed by the formulas
\begin{equation}
\begin{split}
t_{1} =(1 + \cos \theta ) \; \varphi (\chi) \; , \qquad  t_{2} =(1
- \cos \theta ) \; \varphi^{*} (\chi) \;\; ,
\\
\varphi (\chi) = \sin \chi
 \exp\;  [ i({\pi \over 2} - \chi)] \; .
 \end{split}
\label{2.9}
\end{equation}

From (\ref{2.9}) one can  derive the relationship between $t_{1}$ and
$t_{2}$
\begin{eqnarray}
 {t_{1} \over t_{2}^{*} }  =  {t_{1}^{*} \over t_{2} }  =
  - { t_{1} (1 - t_{2} ) \over   t_{2} (1 - t_{1} ) } \; , \;  t_{1}t_{2} = (t_{1} t_{2})^{*} \; ,
\label{2.10a}
\end{eqnarray}
\noindent which are equivalent to
\begin{eqnarray}
t_{1}^{*} = - t_{1}  \; { 1 -t_{2} \over 1 - t_{1}}\; , \qquad
t_{2}^{*} = - t_{2}  \; { 1 -t_{1} \over 1 - t_{2}}\; ,
\label{2.10b}
\end{eqnarray}
\noindent
 or
\begin{equation}
\begin{split}
(1-t_{2}) = (1 -t_{1})\; \left ( - {t_{1} \over t_{1}^{*} } \right
) \; , \qquad
 (1-t_{1}) = (1 -t_{2})\; \left ( - {t_{2} \over
t_{2}^{*} } \right ) \; .
\end{split}
\label{2.10c}
\end{equation}
\noindent
 Its existence may  evidently  be referred  to the real nature of the space $S_{3}$.
In particular, one consequence is: if  $t_{1} \rightarrow 1-0$
then  $t_{2} \rightarrow 1+0$, and so on.

In the following, so defined coordinates
 $(t_{1},t_{2},\phi)$ are called parabolic  coordinates on the sphere $S_{3}$.
 In the limit of the flat space, they  reduce to  the ordinary parabolic coordinates
\begin{eqnarray}
\lim_{\rho \rightarrow \infty } \; (-i \rho t_{1}) = \xi \; ,
\;  \lim_{\rho \rightarrow \infty } \; (-i \rho t_{2}) = -\eta
\;\; . \label{2.11}
\end{eqnarray}

Now, we transform  the metrics of the space $S_{3}$ in spherical
coordinates
\begin{eqnarray}
dl^{2} = -  \;  d \chi ^{2} - \sin ^{2} \chi \; (d \theta ^{2} +
\sin ^{2} \theta \; d\phi ^{2})   \; ,
\nonumber
\end{eqnarray}
\noindent to complex parabolic    $t_{1},t_{2},\phi$. As a first
step, with the help of
\begin{eqnarray}
\sin^{2} \theta =
t_{1} \; t_{2} \; {1 + q^{2} \over q^{2}} \; ,\;
 \sin^{2}
\chi = {q^{2} \over 1 + q^{2} } \; ,
\nonumber
\end{eqnarray}
\noindent we  obtain
\begin{eqnarray}
\sin^{2} \chi \; \sin^{2} \theta \; d \phi^{2} = t_{1}\; t_{2}\; d
\phi^{2} \; .
\label{2.12a}
\end{eqnarray}
\noindent As  a second step,   we  have
\begin{eqnarray}
( d \theta) ^{2} = { 1 \over \sin^{2} \theta } \; (d \; \cos
\theta )^{2} = {1 \over t_{1} t_{2} } \; {q^{2} \over 1 + q^{2}}
\; [ d  ( {t_{1} + t_{2} - 2 t_{1}t_{2} \over t_{1} - t_{2} }  )
]^{2} \; , \nonumber
\end{eqnarray}

\noindent   and further
\begin{eqnarray}
\sin^{2} \chi \; (d \theta)^{2} = {q^{4} \over (1+q^{2})^{2}}\; {4
\over t_{1} t_{2} (t_{1} - t_{2})^{4}}\;
[\; t_{2}(t_{2} - 1) \;
dt_{1} \;  -   \; t_{1}(t_{1} - 1) \; dt_{2}\; ]^{2} \; ,
\nonumber
\end{eqnarray}
\noindent or
\begin{eqnarray}
\sin^{2} \chi \; (d \theta)^{2} = {1 \over 4(1- t_{1})^{2} (1-
t_{2})^{2} t_{1} t_{2} }\;
 [\; t_{2}\; (t_{2} - 1) \; dt_{1} \; -
\;  t_{1} \; (t_{1} - 1) \; dt_{2}\; ]^{2} \; .
\label{2.12b}
\end{eqnarray}
\noindent Taking into account relations
\begin{eqnarray}
{i \; d \chi \over \cos^{2} \chi } =
dt_{1}  {2(1 -t_{2}) \over  (2 - t_{1} - t_{2})^{2} }  -
dt_{2}  {2(1 -t_{1}) \over  ( 2 - t_{1} - t_{2})^{2} }
\nonumber
\end{eqnarray}
\noindent and
\begin{eqnarray}
\cos^{2} \chi = { 1 \over 1 + \mbox{tan}^{2} \chi } = { (2 - t_{1}
- t_{2})^{2} \over 4(1- t_{1}) (1- t_{2})} \; ,
\nonumber
\end{eqnarray}
\noindent we get
\begin{eqnarray}
(d \chi)^{2} = - { [\;
(1-t_{2}) dt_{1} - (1 -t_{1}) dt_{2}\; ]^{2} \over  4(1- t_{1})^{2} (1- t_{2})^{2} }\;   \; .
\label{2.12c}
\end{eqnarray}
\noindent Therefore, for the metrics in parabolic  coordinates in
$S_{3}$
 we have arrived at  the form
\begin{eqnarray}
dl^{2} =  \; {t_{2} - t_{1} \over 4 t_{1} (1 - t_{1})^{2}} \;
dt_{1}^{2}
   +{t_{1} - t_{2} \over 4 t_{2} (1 - t_{2})^{2}} \;
dt_{2}^{2}  \; + \; t_{1} t_{2} d \phi^{2} \;   .
\label{2.13}
\end{eqnarray}
\noindent Formally, this formula differs from  its counterpart in
the space $H_{3}$ only by presence of $(-1)$ in  the expression
for $dl^{2}$.

The more clarity in the complex  coordinates  may be achieved if
one determines them in terms of  Cartesian coordinates
\begin{equation}
\begin{split}
i\; y_{1}= \sqrt{-t_{1} t_{2}}\; \cos \phi \; , \qquad
 i \; y_{2}=
\sqrt{-t_{1}t_{2}}\; \sin \phi \; ,
\\
i \; y_{3}= {t_{1} + t_{2} -2t_{1}t_{2} \over 2
\sqrt{(1-t_{1})(1-t_{2})}}\; , \qquad  y_{0} ={2 -t_{1} - t_{2}
\over 2 \sqrt{(1-t_{1})(1-t_{2})}}\; ;
\end{split}\label{2.14a}
\end{equation}
\noindent they evidently  obey  an  identity (below the notation  $y =\sqrt{y_{1}^{2}+y_{2}^{2} +y_{3}^{2}}$ is used)
\begin{eqnarray}
y_{1}^{2}+ y_{2}^{2}+  y_{3}^{2}+  y_{0}^{2}
 = (y+iy_{0})(y-iy_{0}) = 1\; .
\nonumber
\end{eqnarray}
\noindent Inverse to (\ref{2.14a})
 formulas are
\begin{equation}
\begin{split}
 t_{1} = (y +y_{3}) \; (y+iy_{0}) =a \; e^{i\alpha} \; ,
\\
  t_{2} = (y -y_{3})(y-iy_{0}) = b \; e^{-i\alpha} \;  ,
\\
  t_{1}t_{2} = y_{1}^{2} + y_{2}^{2} \;  ,\;\;
  \;\mbox{tan}\; \phi = {y_{2} \over y_{1}} \; .
\label{2.14b}
\end{split}
\end{equation}
\noindent Two complex coordinates are detailed by
\begin{equation}
\begin{split}
a = y+y_{3}\; , \; b = y -y_{3} \; , \; ab = y_{1}^{2} +
y_{2}^{2} \; ,
\\
\cos \alpha = y\; , \qquad \sin \alpha = y_{0} \; ;
\end{split}
\label{2.15}
\end{equation}

\noindent from this it follows
\begin{eqnarray}
\cos \alpha = y = {a+b \over 2} \; , \qquad {a-b \over 2} = y_{3}
\; .
\nonumber
\end{eqnarray}

The domain for variables  $(a,b)$ can ne illustrated by
the Fig. 1.

\vspace{+35mm} \unitlength=0.6mm
\begin{picture}(160,40)(-65,0)
\special{em:linewidth 0.4pt} \linethickness{0.4pt}

\put(+50,+70){$\alpha \in [\; - {\pi \over 2 }, + {\pi \over 2} \;
] $}

\put(0,0){\vector(+1,0){80}}   \put(+85,-5){$a$} \put(+58,-10){2}
\put(0,0){\vector(0,+1){80}}  \put(-7,+85){$b$}  \put(-10,57){2}

\put(60,0){\line(-1,+1){60}}

\put(0,0){\line(+1,+1){30}}

\put(+30,+30){\circle*{2}}

\put(+20,+10){$y_{3} > 0$}

\put(+7,+25){$y_{3}< 0$}

\end{picture}

\vspace{5mm}

\begin{center}
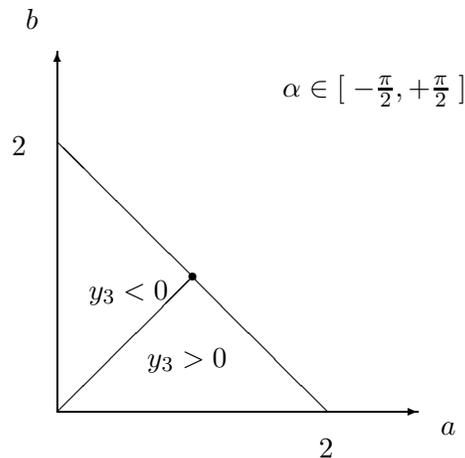
{\bf Fig. 1.  The domain  for  $(a,b)$.}
\end{center}
\noindent Note that associated  points in the sphere   $(y_{0},
y_{k})$   and   $( -y_{0}, -y_{k})$ are parameterized according to
\begin{equation}
\begin{split}
(+y_{0}, +y_{k})  \;\; \Longrightarrow  \;\; (t_{1}, \;  t_{2}, \;
\phi  ) \;,
\\
 (-y_{0}, -y_{k}) \;\;   \Longrightarrow  \;\;  (t_{2}, t_{1}, \phi +
\pi) \; ,
\label{2.16a}
\end{split}
\end{equation}
\noindent or
\begin{equation}
\begin{split}
(+y_{0}, +y_{k})  \;\; \Longrightarrow  \;\; (a, b, \phi  ) \;
,
\\
 (-y_{0}, -y_{k}) \;\;  \Longrightarrow \;\;  (b, a, \phi +
\pi) \; .
\label{2.16b}
\end{split}
\end{equation}

Let us describe peculiarities of parametrization of $S_{3}$ by the
variables
 $(a,b, \alpha, \phi)$.  The first that is a closed line
\begin{eqnarray}
y_{1}=0 \; , \; y_{2}= 0\; , \; y_{3}+ y_{0}^{2}= 1 \; ;
\label{2.17a}
\end{eqnarray}
\noindent  in this case, $ab=0$, and to this line there
corresponds  only a part of  the boundary in Fig. 1, $a=0$ and
$b=0$ (and the coordinate  $\phi$ is  "dumb")
\begin{equation}
\begin{split}
y_{3}> 0 \; , \;   t_{1} = 2y_{3} (y_{3}+iy_{0}) \; ,
\qquad
t_{2} = 0 \; , \; \alpha \in [\; - {\pi \over 2 }, + {\pi
\over 2} \; ] \; ;
\\
y_{3}<  0 \; , \;   t_{1} = 0  \; , \;\;
t_{2} = -2y_{3}
(y_{3}-i y_{0}) \; ,\;\;
 \alpha \in [\; - {\pi \over 2 }, +
{\pi \over 2} \; ] \; .
\label{2.17b}
\end{split}
\end{equation}

Now, let us consider  another closed line
\begin{eqnarray}
y_{3}=0 \; , \; y_{0}= 0\; , \; y_{1}^{2}+ y_{2}^{2}= 1 \;
;
\label{2.18a}
\end{eqnarray}
\noindent characterized by
\begin{equation}
\begin{split}
a =b = \sqrt{y_{1}^{2} + y_{2}^{2}} = 1\; ,
\qquad
 e^{\pm i\alpha}  = 1
\pm i \; 0 \; , \; a+b =2\;  .
\label{2.18b}
\end{split}
\end{equation}
\noindent To this line there correspond a single point  $(1,1)$ on
the  boundary (see Fig. 1), and the coordinate $\phi$ is not
"dump"  now
\begin{eqnarray}
y_{1} = \cos \phi  \; , \qquad  y_{2} = \sin \phi \; .
\label{2.18c}
\end{eqnarray}

There exists one other peculiar region to which there corresponds
the line  $a+b = 2$  on the  boundary (see Fig. 1). Indeed, let
\begin{eqnarray}
y_{0} =0 \;  ,  \qquad y_{1}^{2} + y_{2}^{2} + y_{3}^{2} =1 \; ,
\label{2.19a}
\end{eqnarray}
\noindent  this sphere is parameterized in accordance with
\begin{equation}
\begin{split}
\sin \alpha =0 \; , \;\; \cos \alpha =  1  \qquad  \Longrightarrow \qquad
a+b = 2\; ,\; \phi \in [ 0, 2 \pi ]\; ,
\\
y_{1} =  + \sqrt{a(2-a)}\; \cos \phi \; ,
\qquad
 y_{2} = +
\sqrt{a(2-a)}\; \sin \phi \; , \; y_{3} = a-1  .
\label{2.19b}
\end{split}
\end{equation}
\noindent Here one can introduce the variable  $a-1 =  \cos
\theta$, then
\begin{eqnarray}
y_{1} =  + \sin \theta \cos \phi \; ,
\qquad
 y_{2} = + \sin \theta
\; \sin \phi \; , \; y_{3} = \cos \theta  \; .
\nonumber
\end{eqnarray}

\section{ Separation of variables in parabolic coordinates,
 the models $H_{3}$ and  $S_{3}$
}

Now let us turn to a Coulomb problem.
 General expression for  Schr\"{o}dinger Hamiltonian
\begin{eqnarray}
H =  - {1 \over 2} {1 \over \sqrt{g}} {\partial \over \partial
x^{\alpha}}\; \sqrt{g} g^{\alpha \beta}{\partial \over
\partial x^{\beta}} -  {e \over q} \;  ,
\qquad
\sqrt{g} =  i \; {t_{1} - t_{2} \over 4 (1 - t_{1}) (1 -t_{2}) }
\; ,
\label{3.1}
\end{eqnarray}
\noindent gives

 \underline{$S_{3}$}
\begin{eqnarray}
H =   2  {1 - t_{1} \over t_{1} - t_{2} }  {\partial \over
\partial t_{1}} t_{1}(1 - t_{1}) {\partial \over \partial t_{1}}
 +
  2  {1 - t_{2} \over t_{2} - t_{1} }  {\partial \over
\partial t_{2}} t_{2}(1 - t_{2}) {\partial \over \partial t_{2}}
\nonumber
\\
 -    {1 \over 2 t_{1} t_{2} }\;
{\partial^{2} \over \partial \phi^{2}} \; - \;
 i e \; {2 - t_{1} -t_{2}  \over  t_{1} -t_{2} }   \; .
\label{3.1c}
\end{eqnarray}

\noindent  Transition to the Hamiltonian in the  model $H_{3}$ is
achieved by  performing two formal  replacements:
  $e \;\; \Longrightarrow \;\; ie$, and   $H \;\; \Longrightarrow \;\; -H$.
Thus  we obtain
\begin{widetext}
 \underline{$H_{3}$}
\begin{eqnarray}
H =   -2 \; {1 - t_{1} \over t_{1} - t_{2} } \; {\partial \over
\partial t_{1}} t_{1}(1 - t_{1}) {\partial \over \partial t_{1}}
  -
  2   {1 - t_{2} \over t_{2} - t_{1} } \; {\partial \over
\partial t_{2}} t_{2}(1 - t_{2}) {\partial \over \partial t_{2}}
  \nonumber\\
  + {1 \over 2 t_{1} t_{2} }\;
{\partial^{2} \over \partial \phi^{2}} \; - \;
  e \; {2 - t_{1} -t_{2}  \over  t_{1} -t_{2} } \;  \; .
\label{3.1d}
\end{eqnarray}
\end{widetext}

Now, let us separate the variables in the Schr\"{o}dinger equation
in coordinates  $(t_{1}, t_{2},\phi)$  -- first let us specify the
case of the model $S_{3}$
\begin{eqnarray}
\Psi (t_{1}, t_{2},\phi) = f_{1}(t_{1})\; f_{2}(t_{2})\;e^{im\phi}
\; ; \label{3.2}
\end{eqnarray}
\noindent from  $H\; \Psi = \epsilon \Psi $ it follows
\begin{widetext}
\begin{eqnarray}
f_{2} \; {2(1 - t_{1}) \over t_{1} - t_{2} } \; {d \over dt_{1}}\;
t_{1}(1 -t_{1})\; {d \over dt_{1}}\; f_{1} \; + f_{1} \; {2(1 -
t_{2}) \over t_{2} - t_{1} } \; {d \over dt_{2}}\; t_{2}(1
-t_{2})\; {d \over dt_{2}}\; f_{2} \;
\nonumber
\\
+{ m^{2} \over 2 t_{1} t_{2}}\; f_{1} f_{2} \; - \; ie \; {2 -
t_{1} -t_{2}  \over  t_{1} -t_{2} } \; f_{1} f_{2}  = \epsilon \;
f_{1} f_{2} \; .
\label{3.3}
\end{eqnarray}
\noindent and then multiplying (\ref{3.3})  by  $(t_{1} - t_{2})/2 \;
f_{1} f_{2} \;$,  one derives
\begin{eqnarray}
{1 \over f_{1}} \;(1 -t_{1}) \; {d \over dt_{1}} \; t_{1}(1
-t_{1})\;{d\over dt_{1}}\; f_{1} \; - {m^{2} \over 4 t_{1}} \; +
{ie\over 2} \; t_{1} \; - \; {\epsilon \over 2}\; t_{1}
\nonumber
\\
- \; {1 \over f_{2}} \;(1 -t_{2}) \; {d \over dt_{2}} \; t_{2}(1
-t_{2})\;{d\over dt_{2}}\; f_{2} \; + {m^{2} \over 4 t_{2}} \; +
{ie \over 2} \; t_{2} \; + \; {\epsilon \over 2}  t_{2}   +
(k_{1} - k_{2}) = 0
 \label{3.4a}
\end{eqnarray}
\end{widetext}
\noindent  where two  separation constants  $k_{1}$ and  $k_{2}$
obey an additional constraint
\begin{eqnarray}
k_{1} - k_{2}  = -i\; e \;    .
\label{3.4b}
\end{eqnarray}

Thus, the  problem in  $S_{3}$ consists in solving the system

 \underline{$S_{3}$}
\begin{widetext}
\begin{equation}
\begin{split}
(1 -t_{1}) \; {d \over dt_{1}} \; t_{1}(1 -t_{1})\;{d\over
dt_{1}}\; f_{1} \;  +  \;   ( \; {ie - \epsilon \over 2}\; t_{1}
\; - \; {m^{2} \over 4 t_{1}} \; + \; k_{1} \;) f_{1} = 0 \; ,
\\
(1 -t_{2}) \; {d \over dt_{2}} \; t_{2}(1 -t_{2})\;{d\over
dt_{2}}\; f_{2} \;   + \;  ( \; {- ie - \epsilon \over 2}\; t_{2}
\; - \; {m^{2} \over 4 t_{1}} \; + \; k_{2} \; ) f_{2} = 0 \; .
\label{3.5}
\end{split}
\end{equation}
\noindent In the model $H_{3}$ we obtain
 (separation constants obey  the identity $k_{1} - k_{2}  = e $)

\underline{$H_{3}$}
\begin{equation}
\begin{split}
(1 - t_{1}) \; {d \over dt_{1}} \; t_{1}(1 - t_{1})\;{d \over
dt_{1}}\; f_{1} \;  +  \;   ( \; {-e + \epsilon \over 2}\; t_{1}
\; - \; {m^{2} \over 4 t_{1}} \; + \; k_{1} \; ) f_{1} = 0 \; ,
\\
(1 - t_{2}) \; {d \over dt_{2}} \; t_{2}(1 - t_{2})\;{d\over
dt_{2}}\; f_{2} \;   + \;   ( \; { e + \epsilon \over 2}\; t_{2}
\; - \; {m^{2} \over 4 t_{1}} \; + \; k_{2} \; ) f_{2} = 0 \; .
\label{3.6}
\end{split}
\end{equation}
\noindent Solutions  of  (\ref{3.5}) and  (\ref{3.6}) are searched in the
form
\begin{eqnarray}
f_{1} = t_{1}^{a_{1}} \; (1 - t_{1})^{b_{1}} \; S_{1}(t_{1}) \; ,
\qquad
  f_{2} = t_{2}^{a_{2}} \; (1 - t_{2})^{b_{2}} \;
S_{2}(t_{2}) \;\; .
\label{3.7}
\end{eqnarray}
\noindent It suffice to consider the case  (\ref{3.5}); transition
from $S_{3}$ to  $H_{3}$ is realized trough the changes
\begin{eqnarray}
S_{3} \Longrightarrow H_{3}\; ,  \qquad \epsilon \Longrightarrow -
\epsilon \; , \qquad  e \Longrightarrow +i e\; .
\nonumber
\end{eqnarray}
\noindent The first equation in (\ref{3.5}) gives (to obtain analogous
result for $f_{2}$, is suffices to change  the index   $1$ by  $2$
and the parameter  $e $  by $-e$)
\begin{eqnarray}
t_{1} (1 - t_{1}) \; S''_{1} + S'_{1} \; [ \; 2a (1 -t_{1})
2bt_{1} + (1 - 2t_{1}) \;  ]
\nonumber
\\
  -\left [ \;  a_{1}(a_{1}-1) \left({1\over t_{1}} - 1\right) - 2a_{1}b_{1}
+ b_{1}(b_{1}-1)\left({1\over t_{1}} - 1\right) + a_{1}\left({1\over t_{1}} - 2\right) -
b_{1} \left(2 - {1 \over 1 - t_{1}}\right) \; \right.
\nonumber
\\
\left. +   {ie - \epsilon \over 2}\left({1 \over 1 -t_{1} } -1\right) -
{m^{2} \over 4} \left({1 \over t_{1}} + {1 \over 1 - t_{1}}\right) + k_{1}{1
\over 1-t_{1}}\;  \right ]\; S_{1} (t_{1}) = 0 \; .
\label{3.8}
\end{eqnarray}
\noindent   Terms proportional to $t_{1}^{-1}$ and  $(1 -
t_{1})^{-1}$ can be  eliminated by imposing additional restriction:
\begin{eqnarray}
a^{2}_{1} - {m^{2} \over 4} = 0 \; , \qquad   b^{2}_{1} + {ie -
\epsilon \over 2} -{m^{2} \over 4} + k_{1} = 0 \; ;
\label{3.9}
\end{eqnarray}
\noindent which results in
\begin{eqnarray}
t_{1} (1 - t_{1})  S''_{1} + S'_{1}  [ (2a_{1} + 1)  -
(2a_{1} +  2b_{1} + 2 ) t_{1} ]
\nonumber
\\
-
\left[ a_{1}(a_{1}+1) + 2a_{1}b_{1} +  b_{1}(b_{1}+1) +  {ie
- \epsilon \over 2}   \right] S_{1} = 0  .
\nonumber
 \label{3.10}
 \end{eqnarray}
\noindent This means that   $S(t_{1})$ is expressed in terms of
hypergeometric functions
$
S_{1}(t_{1}) = F(\alpha_{1},  \beta_{1},  \gamma _{1};
t_{1})$
 with parameters obeying to
 \begin{eqnarray}
  \gamma_{1} = 2a_{1}+1 \; , \qquad  \alpha_{1} + \beta _{1} + 1 = 2a_{1}  + 2b _{1} + 2 \;\; ,
\nonumber
\\
 \alpha_{1} \beta_{1} =
a_{1}(a_{1}+1) + 2a_{1}b_{1} +  b_{1}(b_{1}+1) +  {ie - \epsilon
\over 2} \; ,
\nonumber
\end{eqnarray}
\noindent from whence it follows
\begin{eqnarray}
\alpha_{1} = a_{1}   +  b_{1}   + {1 \over 2} +  \sqrt{{1\over 4}
+ {\epsilon -ie \over 2}} \; ,
\nonumber
\\
  \beta_{1} = a_{1}   + b
_{1}  + {1\over 2}  - \sqrt{{1\over 4} + {\epsilon -ie \over 2}}
\; ,\;
 \gamma_{1} = 2a_{1} + 1 \; .
\label{3.11b}
\end{eqnarray}

Let us summarize:

for
 $\underline{S_{3}}$
\begin{equation}
\begin{split}
f_{1} = t_{1}^{a_{1}}\; (1 - t_{1})^{b_{1}}\;S_{1} \; , \;
f_{2} = t_{2}^{a_{2} }\; (1 - t_{2})^{b_{2}}\;S_{2} \;  ,
\\
S_{1} = F(\alpha_{1}\; , \; \beta_{1},\;\gamma_{1}; \; t_{1}) \; ,
\qquad  S_{2} = F(\alpha_{2}, \; \beta_{2},\;\gamma_{2}; \; t_{2})
\; ,
\\
a_{1}^{2} = {m^{2} \over 4 }  \; , \qquad  a_{2}^{2} = {m^{2}
\over 4 }  \; ;\qquad b_{1}^{2} = {\epsilon - i e \over 2} +
{m^{2} \over 4} - k_{1}  \; , \qquad  b_{2}^{2} = {\epsilon + i e
\over 2} + {m^{2} \over 4} - k_{2}  \; ,
\\
\alpha_{1} = a_{1} + b_{1} +  {1\over 2} + \sqrt{{1\over 4} +
{\epsilon -ie \over 2}} \; , \qquad \alpha_{2} = a_{2} + b_{2} +
{1\over 2} + \sqrt{{1\over 4} + {\epsilon + i e \over 2}}  \; ,
\\
\beta_{1} = a_{1} + b_{1} +  {1\over 2} - \sqrt{{1\over 4} +
{\epsilon -ie \over 2}}  \; , \qquad  \beta_{2} = a_{2} + b_{2} +
{1\over 2} - \sqrt{{1\over 4} + {\epsilon + i e \over 2}}  \; ,
\\
\gamma_{1} = 2a_{1} + 1  \; ,   \qquad  \gamma_{2} = 2a_{2} + 1
\;, \qquad k_{1}- k_{2} = -ie;
\label{3.12}
\end{split}
\end{equation}

and for $\underline{H_{3}}$
\begin{equation}
\begin{split}
f_{1} = t_{1}^{a_{1}}\; (1 - t_{1})^{b_{1}}\;S_{1} \; , \qquad
f_{2} = t_{2}^{a_{2} }\; (1 - t_{2})^{b_{2}}\;S_{2} \;  ,
\\
S_{1} = F(\alpha_{1}, \; \beta_{1},\;\gamma_{1}; \; t_{1}) \; ,
\qquad  S_{2} = F(\alpha_{2}, \; \beta_{2},\;\gamma_{2}; \; t_{2})
\; ,
\\
a_{1}^{2} = {m^{2} \over 4 }  \; , \qquad  a_{2}^{2} = {m^{2}
\over 4 }  \; , \qquad b_{1}^{2} = {-\epsilon + e \over 2} +
{m^{2} \over 4} - k_{1}  \; , \qquad  b_{2}^{2} = {-\epsilon - e
\over 2} + {m^{2} \over 4} - k_{2}  \; ,
\\
\alpha_{1} = a_{1} + b_{1} +  {1\over 2} + \sqrt{{1\over 4} +
{-\epsilon + e \over 2}} \; , \qquad \alpha_{2} = a_{2} + b_{2} +
{1\over 2} + \sqrt{{1\over 4} + {-\epsilon -  e \over 2}}  \; ,
\\
\beta_{1} = a_{1} + b_{1} +  {1\over 2} - \sqrt{{1\over 4} +
{-\epsilon + e \over 2}}  \; , \qquad  \beta_{2} = a_{2} + b_{2} +
{1\over 2} - \sqrt{{1\over 4} + {-\epsilon - e \over 2}}  \; ,
\\
\gamma_{1} = 2a_{1} + 1  \; ,   \qquad  \gamma_{2} = 2a_{2} + 1 \;
, \qquad k_{1}- k_{2} = e \;  .
\label{3.13}
\end{split}
\end{equation}
\end{widetext}

\section{ The hydrogen atom in $H_{3}$, bound states
}

The task consists in separating from all solutions found in
Section {\bf IV}
 those  which describe possible bound states.
To treat this problem we will need some additional details in
parametrization  of the model $H_{3}$  by
 coordinates
$t_{1}, t_{2}, \phi$ (see (\ref{2.2}))
\begin{widetext}
\begin{eqnarray}
x_{1} = \sqrt{- t_{1} t_{2} } \cos \phi \; ,
\qquad
 x_{2} =
\sqrt{- t_{1} t_{2} } \sin \phi \; ,
\nonumber
\\
x_{3} = { t_{1} + t_{2} - 2t_{1} t_{2} \over \sqrt{ (1- t_{1})((1-
t_{2})}} \; ,
\qquad  x_{0} = {2 - t_{1} - t_{2} \over \sqrt{ (1-
t_{1})((1- t_{2})}} \; ,
\label{4.2a}
\\
t_{1} = {x_{3} + x \over x_{0} + x } = {q_{3} + q \over 1 + q } \;
, \; 0 \leq  t_{1} <  1 \;  ,
\qquad t_{2} = {x_{3} - x \over x_{0} - x } = { q_{3} - q \over 1 - q }
\; ,
\nonumber
\\
 - \infty \leq t_{2} \leq 0 \; ,
\qquad
\mbox{tan}\; \phi = {x_{2} \over  x_{1}} =  {q_{2} \over  q_{1}}
\; , \qquad \phi \in [ 0 , 2 \pi ] \; .
\label{4.2b}
\end{eqnarray}
\noindent Besides, we will use relations
\begin{eqnarray}
q_{1} = { \sqrt{ - t_{1}  t_{2}} \; \sqrt{ (1- t_{1}) \; (1-t_{2})
} \over 2 -t_{1} - t_{2} } \; \cos \phi \; ,
\nonumber
\\ q_{2} = {
\sqrt{ - t_{1}  t_{2}} \; \sqrt{ (1- t_{1}) \; (1-t_{2}) } \over 2
-t_{1} - t_{2} } \; \sin \phi \; ,
\;
q_{3} = { t_{1} + t_{2} - 2t_{1} t_{2} \over  2 - t_{1} - t_{2}}
\; .
\nonumber
\label{4.2c}
\end{eqnarray}
\end{widetext}
\noindent Note that the origin is characterized by
\begin{eqnarray}
t_{1}=0, \; t_{2}=0  \Longleftrightarrow   q_{1}=0\; ,
\;\; q_{2}=0\; , \;\; q_{3}=0 \; ;
\nonumber
\end{eqnarray}
\noindent and the whole   boundary for the domain $G(t_{1},t_{2})$
is detailed as follows  (see Fig. 2)
\begin{eqnarray}
(a) \qquad t_{1}= 0 \; , \;\; t_{2} \in (- \infty , 0 ]
\Longrightarrow
\nonumber
\\
 q_{1}=0 \;, \; q_{2}= 0 \; ,\;  q_{3} = - q=
{  t_{2}  \over  2 -  t_{2}} \; ,
\nonumber
\\
 t_{2} \rightarrow  - \infty \; , \; q_{3} = -
q \rightarrow -1 \; ;
\nonumber
\label{4.3a}
\\
(b) \qquad t_{1}  \rightarrow + 1 \;  \; , \; t_{2} \in (- \infty
, 0 ]  \Longrightarrow
\nonumber
\\
 q_{1} = 0 \; , \;\; q_{2} =
0\; , \;\; q_{3} \rightarrow  { 1  -   t_{2} \over  1 - t_{2}}  =
+1 \; ,
\nonumber
\label{4.3b}
\end{eqnarray}
\noindent
so the boundary (b) parameterizes  one single point;
\begin{eqnarray}
(c) \qquad t_{1} \in [0, +1 )\; ,\;\; t_{2} = 0 \qquad
\Longrightarrow
\nonumber
\\
q_{1}=0\;, \;\; q_{2} = 0 \; , \;\;
 q_{3} = + q = {t_{1} \over 2 - t_{1} } \; ;
\nonumber
\label{4.3c}
\\
(d)\qquad t_{1} \in [0, +1 )\; ,\;\; t_{2} \rightarrow  -
\infty  \Longrightarrow
\nonumber
\\
 q_{1}=
\sqrt{t_{1}(1-t_{1})}\; \cos  \phi \;,
\nonumber
\\
q_{2}= \sqrt{t_{1}(1-t_{1})}\; \sin \phi \;, \qquad q_{3} = 2t_{1}
-1 \; .
\nonumber
\label{4.3d}
\end{eqnarray}
\noindent
It is ellipsoid  $4(q_{1}^{2} + q_{2}^{2} ) + q_{3}^{2} = 1 $
passing through two points
\begin{eqnarray}
 t_{1} \rightarrow 1 \;  q_{1} \rightarrow 0 \;
, \; q_{2} \rightarrow 0 \;  , \; q_{3} \rightarrow + 1 \;
,
\nonumber
\\
t_{1} \rightarrow 0 , \;   q_{1} \rightarrow 0 \;
, \;  q_{2} \rightarrow 0 \;  , \;  q_{3} \rightarrow - 1 \;
. \label{4.3e}
\end{eqnarray}

The structure of the boundary  may be illustrated  by Fig. 2.
Some clarity can be added  with the help of the   inverse formulas
(let $q_{i} = q \; n_{i}$):
\begin{eqnarray}
t_{1} =   {q  (n_{3} +1)\over 1 + q}\; , \; t_{2} =
  {q (n_{3}-1) \over 1 - q}\; ,\;
 \mbox{tan}\; \phi = {n_{2}
\over n_{1}} \; ;
\nonumber
\label{4.4a}
\end{eqnarray}
\noindent from whence it follows
\begin{eqnarray}
n_{3} = +1 \; , \qquad  t_{1} = 0 \;, \qquad   t_{2} = {2q \over
q-1}\; ,
\nonumber
\\
n_{3} = -1 \; , \qquad  t_{1} =  {2q \over q+1}\; ,  \qquad  t_{2}
= 0 \; .
\nonumber
\end{eqnarray}

\vspace{+10mm}
\unitlength=0.65mm
\begin{picture}(160,40)(-90,0)
\special{em:linewidth 0.4pt} \linethickness{0.4pt}

\put(0,0){\line(-1,0){50}}  \put(0,0){\vector(+1,0){20}}
\put(+22,-5){$t_{2}$}

\put(0,0){\vector(0,+1){50}}  \put(+2,+50){$t_{1}$}
  \put(0,+20){\line(-1,0){50}}
\put(0,+21){\line(-1,0){50}}

\put(0,0){\circle*{2}}  \put(70,10){\circle*{2}}


\put(+70,+10){\circle{20}} \put(+70,+10){\vector(+1,0){30}}

\put(+70,+10){\vector(0,+1){30}}

\put(+70,+10){\vector(-1,-1){15}}

\put(-65,+8){$(d)$}  \put(+2,+8){$(c)$}

\put(-30,-8){$(a)$}  \put(-30,+22){$(b)$}

\put(+80,+40){$q_{i}$-space}  \put(-67,+40){$(t_{1},t_{2},\phi)$-space}

\put(-55,+1){\line(0,+1){2}} \put(-55,+4){\line(0,+1){2}}
\put(-55,+7){\line(0,+1){2}} \put(-55,+10){\line(0,+1){2}}
\put(-55,+13){\line(0,+1){2}} \put(-55,+16){\line(0,+1){2}}

\end{picture}


\vspace{5mm}

\begin{center}
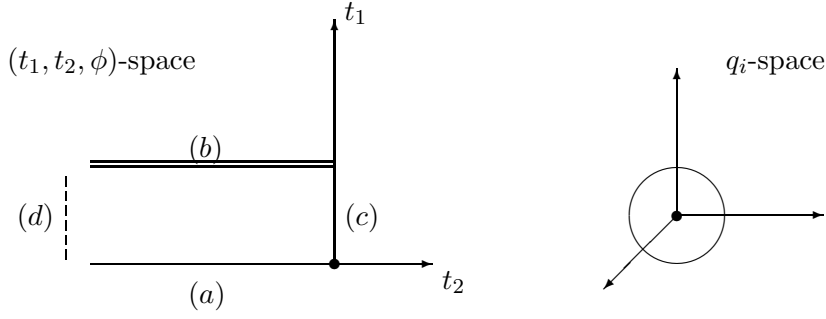
{\bf Fig. 2. Parabolic coordinates in  $H_{3}$}

\end{center}

\noindent In particular, at $n_{3} \neq \pm 1$ the boundary  $q  =
(1 - \Delta) , \; \Delta \rightarrow +0 $ is parameterized
according to
\begin{eqnarray}
t_{1} \rightarrow { ( 1 + n_{3} ) \over 2} \; , \; t_{2}
\rightarrow  -  { (1 - n_{3} )  \over  \Delta } \; , \;
\mbox{tan}\; \phi = {n_{2} \over n_{1}} \; .
\nonumber
\label{4.4b}
\end{eqnarray}

 Now we are ready to construct the bound states
solutions. In the firs place, note that to have solutions
vanishing in the  origin $q_{i}=0$,  we must take $a_{1}$ and
$a_{2}$ positive
\begin{eqnarray}
 a_{1} = + { \mid m \mid \over 2} \; , \qquad
a_{2} = + { \mid m \mid \over 2} \; .
\label{4.5a}
\end{eqnarray}

\noindent To have solution single-valued and continuous in the
region $ q_{3}  \rightarrow  +1  \; ,  \; q_{1}=0 \; ,  \;
q_{2} = 0$ (the  boundary  (b) in the Fig. 2), we  must takes
positive  $b_{1}$ and negative $b_{2}$ :
\begin{eqnarray}
b_{1}  = + \sqrt{{-\epsilon + e \over 2} + {m^{2} \over 4} - k_{1}
}   > 0  \;,
\nonumber
\\
 b_{2}  = - \sqrt{{-\epsilon + -e \over 2} +
{m^{2} \over 4} - k_{1} }  <  0  \; ;
\label{4.5b}
\end{eqnarray}

\noindent besides one should check that the total negative power
$(a_{2} + b_{2})$ compensates a positive power $n_{2}$ of the
quantum number
 of the main term of  a  polynomial at infinity
\begin{eqnarray}
a_{2} + b_{2} +  n_{2}  < 0 \; \; .
\label{4.5c}
\end{eqnarray}

\noindent Additionally we assume positiveness of two expressions
under the square roots in (\ref{4.5b}).

To obtain a polynomial in the variable $t_{1}$, we require
\begin{eqnarray}
\beta_{1} =   - n_{1}\; , \qquad n_{1} = 0, 1, 2, 3, ...
\label{4.6a}
\end{eqnarray}

\noindent In turn, the same in variable $t_{2}$ can be reached by
\begin{eqnarray}
\beta_{2} =   - n_{2}\; , \qquad n_{2} = 0, 1, 2, 3, ...
\label{4.6b}
\end{eqnarray}

Equations  (\ref{4.6a}) and  (\ref{4.6b})   will give  (let  $N_{1} = 2 n_{1} + \mid m
\mid + 1$  and  $ N_{2} = 2 n_{2} + \mid m \mid + 1$)
\begin{eqnarray}
  \sqrt{ 2 (e - \epsilon ) + m^{2} - 4k_{1} } = \sqrt{ 1 + 2 (+ e - \epsilon) }  - N_{1}  ,
\nonumber
\\
  \sqrt{ 2 (-e - \epsilon ) + m^{2} - 4k_{2} } =   N_{2}  - \sqrt{ 1 + 2 (-e - \epsilon) }    ,
\nonumber
\end{eqnarray}

\noindent Squaring both ones,  after simple
manipulation we get
\begin{eqnarray}
[\; \sqrt{ 1 + 2 (+e - \epsilon) } - N_{1} \; ] ^{2}  =
 [\; N_{2}
-  \sqrt{ 1 + 2 (-e - \epsilon) }  \; ]^{2}  ,
\nonumber
\end{eqnarray}

\noindent so that
\begin{eqnarray}
 \sqrt{ 1 + 2 (+ e - \epsilon) } - N_{1} = N_{2} -  \sqrt{ 1 + 2 (-e - \epsilon) }
\nonumber
\end{eqnarray}

\noindent or
\begin{eqnarray}
 \sqrt{ 1 + 2 (e - \epsilon) }  + \sqrt{ 1 + 2 (-e - \epsilon) }  =2k \; ,
\qquad   k =  {N_{1}  + N_{2} \over 2} = n_{1} + n_{2} + \mid m \mid + 1  .
\nonumber
\label{4.9}
\end{eqnarray}

\noindent Squaring the above equation, we arrive at a quadratic
equation
 (let  $x = 1 -2 \epsilon$)
\begin{eqnarray}
x^{2} - 4e^{2}  = ( 2k^{2} - x)^{2} \; ,
\nonumber
\end{eqnarray}

\noindent with solution
\begin{eqnarray}
\epsilon = - {e^{2} \over 2k^{2} } - {k^{2} - 1 \over 2}\; ,
\qquad
 k = n_{1} + n_{2} + \mid m \mid + 1 = 1, 2, 3, ...
\label{4.10}
\end{eqnarray}

\noindent  The energy levels belong
to the interval
\begin{eqnarray}- \; {e^{2} \over 2} \leq \epsilon \leq ({1\over 2}  - e ) \;, \;  k < \sqrt{e} \; .
\label{4.11}
\end{eqnarray}

It is a matter os simple calculations to find expressions for
 $k_{1}, k_{2}$ and involved parameters. To this end, first
 let us derive simple expressions for roots
\begin{eqnarray}
+ \sqrt{ {1 \over 4} + {e - \epsilon  \over 2} } =+ {1 \over 2}\; (k + {e \over  k} ) \; ,
\qquad
+ \sqrt{ {1 \over 4} + {- e - \epsilon  \over 2} } = + {1 \over 2}\; (k - {e \over  k} ) \;  ;
\label{4.12}
\end{eqnarray}

\noindent then
\begin{eqnarray}
a_{1} + b_{1} + { 1 \over 2} =  -n_{1} + {1 \over 2}\; (k +  {e
\over  k} )   \; , \qquad
a_{2} + b_{2} + { 1 \over 2} = -n_{2} + {1 \over 2}\; (k -  {e
\over  k} )   \; .
\label{4.13}
\end{eqnarray}

\noindent It is easily checked
 (\ref{4.5c}):
\begin{eqnarray}
a_{2} + b_{2} + n_{2} < 0 \qquad   \Longrightarrow
\nonumber
\\
\left ( -n_{2}  - {1 \over 2} + {1 \over 2}\; (k -  {e \over  k} )
\right ) + n_{2} =
-{1\over 2} + {1 \over 2}\; (k -  {e \over  k}
) < 0 \; ;
\label{4.13c}
\end{eqnarray}

\noindent which holds if  (see (\ref{4.11}))
\begin{eqnarray}
k -  {e \over  k} < 0 \;  \qquad \Longleftrightarrow \qquad k  <
\sqrt{e}\; .
\label{4.13d}
\end{eqnarray}

Now, expressions for
 $\alpha_{1}, \alpha_{2}$ are
\begin{eqnarray}
\alpha_{1} = -n_{1} +  (k + {e \over  k} )  = n_{2} +  \mid m \mid
+ 1 + {e \over  k}  \; ,
\nonumber
\\
\alpha_{2} = -n_{2} +  (k -  {e \over  k} ) - N_{1} =
n_{1} +
\mid m \mid + 1 - {e \over  k}  .
\nonumber
\label{4.14}
\end{eqnarray}

And finally, for  $b_{1}, b_{2}$
\begin{eqnarray}
b_{1} =  + {1 \over 2 }\;  \sqrt{   2 (e - \epsilon  )  + m^{2} -
4 k_{1}  }  =
 {1 \over 2} \; [\;
  + \sqrt{ 1 + 2( e - \epsilon ) } - N_{1}  \;   ]
\nonumber
\\
  = {1 \over 2}  \;  \left [\;    (k + {e \over  k} )  - N_{1} \; \right ]  =
  {1 \over 2}  \;  \left [\; + ( n_{2} - n_{1})  + {e \over  k} \; \right ] > 0\; ,
\nonumber
 \label{4.15a}
 \\
 b_{2} =  - {1 \over 2 }\;  \sqrt{ 2 ( - e   -  \epsilon)    +
m^{2}  - 4k_{2} }  =
 {1 \over 2}\; [\;  + \sqrt{ 1  + 2( -e - \epsilon ) }  -   N_{2}   \; ]
\nonumber
\\
= {1 \over 2}  \; \left [ \;    (k  - {e \over  k} )  - N_{2} \;
\right ]  =
{1 \over 2}  \;  \left [\; - ( n_{2} - n_{1}) -  {e
\over  k} \; \right ]  < 0\; ;
\nonumber
\label{4.15b}
\end{eqnarray}

\noindent it must hold  restriction $n_{2} > n_{1}$. Besides, note
that
\begin{eqnarray}
b_{1} + b_{2} = 0 \; .
 \label{4.15c}
 \end{eqnarray}

\noindent Additionally, expressions for
 $\beta_{1} , \beta_{2}$ can be  verified
\begin{eqnarray}
\beta_{1} = {\mid m \mid \over 2} + {1 \over 2} + b_{1} -
\sqrt{{1\over 4} +{ -\epsilon + e \over 2}}
\nonumber
\\
 = {\mid m \mid \over 2} + {1 \over 2} + {1 \over 2}    \left
[    (k +  {e \over  k} )  - N_{1}  \right ]
  - {1 \over 2}\; (k +  {e \over  k} ) =
  { \mid m \mid +1 -N_{1} \over 2} = -n_{1} \; ,
\nonumber
\\
\beta_{2} = {\mid m \mid \over 2} + {1 \over 2} + b_{2} -
\sqrt{{1\over 4} +{- \epsilon - e \over 2}}
\nonumber
\\
 = {\mid m \mid \over 2} + {1 \over 2} + {1 \over 2}    \left
[    (k - {e \over  k} )  - N_{2}  \right ]
  - {1 \over 2}\; (k - {e \over  k} ) =
   { \mid m \mid +1 -N_{2} \over 2} = -n_{2} \; .
\label{4.16}
\end{eqnarray}

Let us find expressions for
 separation constants  $k_{1}, k_{2}$
\begin{eqnarray}
    -2\epsilon  +  2 e   + m^{2} - 4 k_{1}    =
   [\;  ( n_{2} - n_{1}) + {e \over  k} \;  ] ^{2}\; ,
\nonumber
\\
- 2\epsilon  -  2 e  + m^{2}  - 4k_{2}   =
 [\;  ( n_{2} - n_{1})  + {e \over  k} \;  ]^{2} \; ;
\nonumber
\end{eqnarray}

\noindent from whence it follow
\begin{eqnarray}
 4 k_{1} =  (k +  {e \over k} )^{2}  -
 [  ( n_{2} - n_{1}) +  {e \over  k}   ] ^{2} +m^{2} - 1\; ,
\nonumber
\\
  4 k_{2}  =  (k -  {e \over k} )^{2} -
  [  ( n_{2} - n_{1}) +  {e \over  k}   ] ^{2} +m^{2} - 1 \; .
\label{4.18}
\end{eqnarray}

\noindent In particular,  one obtains $ k_{1} - k_{2} = -i e\; . $

\section{The hydrogen atom in space $S_{3}$,  bound states in  complex parabolic coordinates
}

The task consists in separating from all solutions found in
Section {\bf 4} those  which describe possible bound states. We will
need some  details in parametrization  of the model $H_{3}$
\begin{eqnarray}
i \; y_{1}= \sqrt{-t_{1}t_{2}}\; \cos \phi \; , \qquad i \; y_{2}=
\sqrt{-t_{1}t_{2}}\; \sin \phi \; ,
\nonumber
\\
i\; y_{3}= {t_{1} + t_{2} -2t_{1}t_{2} \over 2
\sqrt{(1-t_{1})(1-t_{2})}}\; ,
\qquad
 y_{0} ={2 -t_{1} - t_{2}
\over 2 \sqrt{(1-t_{1})(1-t_{2})}}\; ,
\nonumber
\\
 t_{1} = (y +y_{3}) \; (y+iy_{0}) =a \; e^{i\alpha} \; ,
 \qquad
  t_{2} = (y -y_{3})(y-iy_{0}) = b \; e^{-i\alpha} \;  ,
 \nonumber
 \\
 \mbox{tan}\; \phi = {y_{2} \over y_{1}} \; ,  \qquad  t_{1}t_{2} = y_{1}^{2} + y_{2}^{2} \; .
\label{5.2}
\end{eqnarray}

Let us consider peculiarities of the parametrization with
$(t_{1}, t_{2} ;  \phi) $ - $(a,b, \alpha ;   \phi)$. The
origin is described by
\begin{eqnarray}
t_{1} = 0\; , \;\; t_{2} = 0  \; (a=0\; ,\; b=0) \;\qquad
\Longrightarrow
\qquad
  y_{i}=0 \; , \;\;  y_{0} = +1 \; .
\nonumber
\end{eqnarray}

\noindent The  closed line
 $y_{0}^{2} + y_{3}^{2} = 1$  consists of two parts:
\begin{eqnarray}
t_{1} = 0 \; , \; t_{2} \neq  0 \;  (a =0\; , \; b \neq 0 )
\qquad   \Longrightarrow
\qquad
i y_{3} = { t_{2} \over  2 \sqrt{1 - t_{2} } } \; , \; y_{0} = {
2 - t_{2} \over  2 \sqrt{1 - t_{2} }} \; ;
\label{5.3a}
\end{eqnarray}

\noindent and
\begin{eqnarray}
t_{1} \neq  0 \; , \; t_{2} =  0 , \;   (a \neq 0\; , \; b = 0 )
\qquad  \Longrightarrow \qquad
i y_{3} = { t_{1} \over  2 \sqrt{1 - t_{1} } } \; , \;\; y_{0} = {
2 - t_{1} \over  2 \sqrt{1 - t_{1} }} \;  ;
\label{5.3b}
\end{eqnarray}

\noindent here the coordinate $\phi$ is a  "dump" \hspace{2mm}
one. For another closed  line
\begin{eqnarray}
y_{3}=0 \; , \qquad y_{0}= 0\; , \qquad y_{1}^{2}+ y_{2}^{2}= 1 \;
; \label{5.4a}
\end{eqnarray}

\noindent we  have description
\begin{eqnarray}
a =b = \sqrt{y_{1}^{2} + y_{2}^{2}} = 1\; ,
\;\;
 e^{\pm i\alpha}  = 1
\pm i \; 0 \; ,
\nonumber
\\
 a+b =2\;  ,\;\;
t_{1} = e^{i\alpha} \rightarrow 1 \;, \;\; t_{2} = e^{-\alpha}
\rightarrow 1 \; ,
\nonumber
\\
t_{1} t_{2} \rightarrow  1 \;, \qquad
y_{1}=  \cos \phi \; , \qquad y_{2}=
 \sin \phi \; ,
\nonumber
\\
iy_{3}=   {t_{1} + t_{2} -2   \over 2 \sqrt{ 2 -t_{1} - t_{2} } }
= { i \over 2} \;  \sqrt{2 - t_{1}-t_{2} } =
 {i \over \sqrt{2}} \;
\sqrt{ 1 -  \cos \alpha   } = 0 \; ,
\nonumber
\\
 y_{0} ={2 -t_{1} - t_{2}
\over 2 \sqrt{2 - t_{1} - t_{2} }}= { 1 \over 2} \;  \sqrt{2 -
t_{1}-t_{2} } =
{1 \over \sqrt{2}} \; \sqrt{ 1 -  \cos \alpha   }
= 0 \; .
\label{5.4}
\end{eqnarray}

 This means that to the line (\ref{5.4a}) there corresponds
one single point  $(1,1)$ on the boundary (see Fig. 1), and here
the coordinate  $\phi$ is not "dump". There exists one other
peculiar region to which the boundary   $a+b = 2$  (see Fig. 1)
is referred
\begin{eqnarray}
y_{0} =0 \;  ,  \qquad y_{1}^{2} + y_{2}^{2} + y_{3}^{2} =1 \; ,
\label{5.5a}
\end{eqnarray}

\noindent  at this
\begin{eqnarray}
\sin \alpha =0 \; , \; \cos \alpha =  1 \; , \; t_{1}=a\;,
\;\; t_{2}= b \; \Longrightarrow
\nonumber
\\
 t_{1} + t_{2} = a+b = 2\; ,\; \phi \in [ 0, 2 \pi ]\; ,
\nonumber
\\
y_{1} =  + \sqrt{a(2-a)}\; \cos \phi \; ,
\nonumber
\\ y_{2} = +
\sqrt{a(2-a)}\; \sin \phi \; , \qquad y_{3} = a-1  \; .
\label{5.5b}
\end{eqnarray}

Finally, we must remember on  special additional restriction to
which obey two complex coordinates
\begin{eqnarray}
(1-t_{2}) = (1 -t_{1})\; \left ( - {t_{1} \over t_{1}^{*} } \right
) \; , \qquad
 (1-t_{1}) = (1 -t_{2})\; \left ( - {t_{2} \over
t_{2}^{*} } \right ) \; .
\label{5.6}
\end{eqnarray}

\noindent In particular, these means that if $t_{1} \rightarrow 1
\pm 0$  then  $t_{2} \rightarrow 1 \mp0$, and inversely.

Now let us separate solutions for bound states. To have functions
vanishing on the axis   $y_{1}=0, y_{2}=0$, we must
take   positive $a_{1}$ and  $a_{2}$:
\begin{eqnarray}
a_{1} = + { \mid m \mid \over 2} \; , \qquad a_{2} = + { \mid m
\mid \over 2} \; .
\label{5.7a}
\end{eqnarray}

Let us consider behavior of the functions on the line $y_{1}^{2} +
y_{2}^{2}=1$ -- see (\ref{5.4}). Because here  $t_{1} \rightarrow 1\;
, \; t_{2} \rightarrow 1 $,  to have continuous and finite
solutions we must impose restrictions
\begin{eqnarray}
\mbox{Re}\;  b_{1} > 0 \; , \; b _{1} =  + \; \sqrt{{\epsilon
- i e \over 2} + {m^{2} \over 4} - k_{1}} \; ,
\nonumber
\\
\mbox{Re}\;  b_{2} > 0 \; , \; b _{2} =  + \; \sqrt{{\epsilon
+ i e \over 2} + {m^{2} \over 4} - k_{2}  } \; .
\label{5.7b}
\end{eqnarray}

\noindent In general, instead it is enough to require only
\begin{eqnarray}
\mbox{Re}\; ( b_{1} + b_{2}  ) \geq  0 \;  .
\label{5.7c}
\end{eqnarray}

Assuming
\begin{eqnarray}
\mbox{Re}\;  \left  ( +  \sqrt{  {1\over 4} + {\epsilon - i e
\over 2}  } \right )  > 0 \; , \qquad
 \mbox{Re}\; \left ( + \sqrt{
{1\over 4} + {\epsilon + i e \over 2} } \right  ) > 0 \; ,
\nonumber
\end{eqnarray}

\noindent let us reduce  hypergeometric functions into polynomials
\begin{eqnarray}
\beta_{1} = -n_{1} \;, \qquad n_{1} = 0, 1, 2, ...  ;
\nonumber
\\
\beta_{2} = -n_{2} \;, \qquad n_{2} = 0, 1, 2, ...;
\label{5.8}
\end{eqnarray}

\noindent thereby it is  supposed that there exist such values for
energy and complex  $k_{1},k_{2}$ at which imaginary parts of two
square roots in $\beta_{1}$ and $\beta_{2}$ will cancel out each
other. Eqs. (\ref{5.8}) give
\begin{eqnarray}
\beta_{1} =  {\mid m \mid \over 2}  + \sqrt{{\epsilon - i e \over
2} + {m^{2} \over 4} - k_{1}}
 +
  {1\over 2} - \sqrt{{1\over 4} +
{\epsilon -ie \over 2}}  = -n_{1} \; ,
\nonumber
\\
\beta_{2} =  {\mid m \mid \over 2}  + \sqrt{{\epsilon + i e \over
2} + {m^{2} \over 4} - k_{2}} +
{1\over 2} - \sqrt{{1\over 4} +
{\epsilon + i e \over 2}}  = - n_{2}  \; ,
\label{5.9}
\end{eqnarray}

\noindent or differently
\begin{eqnarray}
  \mid m \mid +   1  + 2n_{1}  + \sqrt{   2(\epsilon - i e )  + m^{2} - 4 k_{1} }
  =
   + \sqrt{ 1 + 2(\epsilon -ie ) } \; ,
\nonumber
\\
  \mid m \mid +
1   + 2n_{2}   + \sqrt{ 2(\epsilon + i e)  + m^{2}  - 4k_{2}}   =
+\sqrt{ 1  + 2(\epsilon + i e) }  \; .
\label{5.10a}
\end{eqnarray}

\noindent Let
\begin{eqnarray}
N_{1} = 2n_{1}  + \mid m \mid  +1 \;, \; N_{2} = 2n_{2}+  \mid
m \mid  +1 \;,
\nonumber
\end{eqnarray}

\noindent then eqs. (\ref{5.10a}) take the form
\begin{eqnarray}
+ \sqrt{   2\epsilon  - 2i e   + m^{2} - 4 k_{1}  }  =
    +
\sqrt{ 1 + 2(\epsilon -ie ) } - N_{1}    \; ,
\nonumber
\\
+ \sqrt{ 2\epsilon  + 2i e  + m^{2}  - 4k_{2} }  =
  + \sqrt{ 1  +
2(\epsilon + i e) }  -   N_{2}    \; .
 \label{5.10b}
 \end{eqnarray}

\noindent Squaring both equations, after simple manipulation we
get (remembering on $k_{1} - k_{2} = -ie$)
\begin{eqnarray}
[  \; +  \sqrt{ 1 + 2(\epsilon -ie ) } - N_{1} \; ]^{2} =
  [ \; +
\sqrt{ 1  + 2(\epsilon + i e) }  -N_{2} \; ] ^{2} \; ;
\nonumber
\end{eqnarray}

\noindent from this two equations follow
\begin{eqnarray}
  \sqrt{ 1 + 2(\epsilon -ie ) } - N_{1}= + [  \sqrt{ 1  + 2(\epsilon + i e) }  -N_{2}  ]\; ,
\nonumber
\\
  \sqrt{ 1 + 2(\epsilon -ie ) } - N_{1}= - [  \sqrt{ 1  +
2(\epsilon + i e) }  -N_{2}  ]\; .
  \nonumber
  \end{eqnarray}

\noindent They give respectively
\begin{eqnarray}
  \sqrt{ 1 + 2(\epsilon -ie ) } - \sqrt{ 1  + 2(\epsilon + i e) } =
   N_{1}  -N_{2} = 2n \; ,
\nonumber
\\
\sqrt{ 1 + 2(\epsilon -ie ) }  +   \sqrt{ 1  + 2(\epsilon + i e) } =
  N_{1} +  N_{2} = 2k \; .
\label{5.11}
\end{eqnarray}

 Note, that the first of them cannot  be valid,
further we will consider only second equation in (\ref{5.11}) -- it
represent a  correct rule for energy quantization. Squaring it we
obtain
\begin{eqnarray}
 (1  + 2\epsilon )   +   \sqrt{ (1 + 2\epsilon)^{2} +4e^{2}} =   2k^{2}  \; .
   \label{5.12}
   \end{eqnarray}

\noindent Its solution is (let $2\epsilon +1=x$)
\begin{eqnarray}
 \epsilon =  - {e^{2} \over 2\; k^{2} } + {k^{2} -1 \over 2} \; ,
 \qquad
  k = n_{1} + n_{2} + \mid m \mid + 1 \; .
 \label{5.13}
 \end{eqnarray}

With the used of the formula for energy levels
\begin{eqnarray}
2\epsilon + 1 = - {e^{2} \over k^{2}}  + k^{2} \; ,
\nonumber
\end{eqnarray}

\noindent one can derives simple expressions for all involved
parameters. First, one finds
\begin{eqnarray}
+ \sqrt{ {1 \over 4} + {\epsilon - ie \over 2} } =
+
 {1 \over 2} \; \sqrt{  k^{2} - {e^{2} \over k^{2} } - ie } = + {1 \over 2}\; (k - i{e \over  k} ) \; ,
\nonumber
\\
+ \sqrt{ {1 \over 4} + {\epsilon + ie \over 2} } =
+
 {1 \over 2} \; \sqrt{  k^{2} - {e^{2} \over k^{2} } + ie } =
 + {1 \over 2}\; (k + i{e \over  k} ) \;  .
\nonumber
\label{5.14}
\end{eqnarray}

\noindent Now expressions for
\begin{eqnarray}
a_{1} + b_{1} + { 1 \over 2} = -n_{1} + {1 \over 2}\; (k - i{e
\over  k} )   \; ,
\nonumber
\\
a_{2} + b_{2} + {1 \over 2} = -n_{2} + {1 \over 2}\; (k - i{e
\over  k} )   \; ,
\nonumber
\label{5.15}
\end{eqnarray}

\noindent and then  $\alpha_{1}, \alpha_{2}$:
\begin{eqnarray}
\alpha_{1} = -n_{1} +  (k - i{e \over  k} )  = n_{2} +  \mid m
\mid + 1 - i{e \over  k}  \; ,
\nonumber
\\
\alpha_{2} = -n_{2} +  (k + i{e \over  k} )= n_{1} +  \mid m \mid
+ 1 + i{e \over  k} \; .
\nonumber
\label{5.16}
\end{eqnarray}

\noindent For  $b_{1}, b_{2}$ we easily  produce
\begin{eqnarray}
b_{1} =  + {1 \over 2 }\;  \sqrt{   2\epsilon  - 2i e   + m^{2} -
4 k_{1}  }  =
  {1 \over 2} \; [\;
  + \sqrt{ 1 + 2(\epsilon -ie ) } - N_{1}  \;   ] =
  {1 \over 2}  \;  \left [\; + ( n_{2} - n_{1}) - i {e \over  k} \; \right ]\; ,
\nonumber
\\
b_{2} =  + {1 \over 2 }\;  \sqrt{ 2\epsilon  + 2i e  + m^{2}  -
4k_{2} }  =
{1 \over 2}\; [\;  + \sqrt{ 1  + 2(\epsilon + i e) }
-   N_{2}   \; ]  =
{1 \over 2}  \;  \left [\; - ( n_{2} - n_{1}) + i {e
\over  k} \; \right ] \; .
\nonumber
\label{5.17b}
\end{eqnarray}

It should be noted that an identity holds
\begin{eqnarray}
b_{1} + b_{2} = 0 \; ;
\label{5.17c}
\end{eqnarray}

\noindent the latter is enough to ensure finiteness of the
solutions at the region
  $t_{1} \rightarrow  1, \; t_{2} \rightarrow 1$.

  It was emphasized  above that it has sense
  to examine continuity properties in accordance with    (see Fig. 1)
\begin{eqnarray}
t_{1} = t_{2} = a+b = 2 \; \Longrightarrow \; (1-t_{1}) =
- (1 - t_{2})\; ;
\nonumber
\end{eqnarray}

\noindent from whence if follows that if  $(1-t_{1})=0$, then  $
(1-t_{2})=0$. In other words, we have no ground to expect
continuity of the following types
\begin{eqnarray}
\Psi (t_{1} \rightarrow 1, t_{2}  ) \;, \qquad  \mbox{or} \qquad
\Psi (t_{1} , t_{2}  \rightarrow 1 ) \; .
\nonumber
\end{eqnarray}

Additionally, one can check expressions for
 $\beta_{1} , \beta_{2}$
\begin{eqnarray}
\beta_{1} = {\mid m \mid \over 2} + {1 \over 2} + b_{1} -
\sqrt{{1\over 4} +{\epsilon -ie \over 2}}=
\nonumber
\\
  {\mid m \mid \over 2} + {1 \over 2} + {1 \over 2}    \left
[\;    (k - i{e \over  k} )  - N_{1} \right ]
  - {1 \over 2} (k - i{e \over  k} ) =
  { \mid m \mid +1 -N_{1} \over 2} = -n_{1}  ,
\nonumber
\\
\beta_{2} = {\mid m \mid \over 2} + {1 \over 2} + b_{2} -
\sqrt{{1\over 4} +{\epsilon +ie \over 2}}=
\nonumber
\\
  {\mid m \mid \over 2} + {1 \over 2} + {1 \over 2}    \left
[    (k + i{e \over  k} )  - N_{2}  \right ]
  - {1 \over 2} (k + i{e \over  k} ) =
   { \mid m \mid +1 -N_{2} \over 2} = -n_{2}.
\nonumber
\label{5.18}
\end{eqnarray}

Finally, it is the matter of simple calculations to specify
$k_{1}$ and $k_{2}$:
\begin{eqnarray}
    2\epsilon  - 2i e   + m^{2} - 4 k_{1}    =
   [\; + ( n_{2} - n_{1}) - i {e \over  k} \;  ] ^{2}\; ,
\nonumber
\\
2\epsilon  + 2i e  + m^{2}  - 4k_{2}   =
 [\; - ( n_{2} - n_{1}) + i {e \over  k} \;  ]^{2} \; ,
\nonumber
\end{eqnarray}

\noindent that is
\begin{eqnarray}
 4 k_{1} =  (k - i {e \over k} )^{2}  -
 [\;  ( n_{2} - n_{1}) - i {e \over  k} \;  ] ^{2} +m^{2} - 1\;,
\nonumber
\\
  4 k_{2}  =  (k + i {e \over k} )^{2} -
  [\;  ( n_{2} - n_{1}) - i {e \over  k} \;  ] ^{2} +m^{2} - 1 \; .
\nonumber
\end{eqnarray}

\noindent In particular, the  identity  $ k_{1} - k_{2} = -i
e$ holds.

\section{
The Runge--Lenz vector  and  parabolic
coordinates }

At separating the variables in Schr\"{o}dinger equation two
constants were introduced $k_{1}$ Ё $k_{2}$; the problem is to
find an operator that is diagonalized on wave functions (31) with
eigenvalues $(k_{1}+k_{2})$
\begin{eqnarray}
\hat{B} \; f_{1} \; f_{2} \; e^{im\phi} = ( k_{1}+k_{2} ) \; f_{1}
\; f_{2} \; e^{im\phi}  \; .
\label{6.1}
\end{eqnarray}

\noindent For the operator$\hat{B}$
one can obtain  the following representation
\begin{widetext}
\begin{eqnarray}
\hat{B} =   -(1 - t_{1}) \; {\partial \over \partial t_{1}} \;
t_{1} (1 - t_{1}) \; {\partial \over \partial t_{1}} \;   - t_{1}
\; {( - H + ie ) \over 2}  - {1 \over 4 t_{1}} \; {\partial^{2}
\over \partial \phi^{2}}
\nonumber
\\
-(1 - t_{2}) \;  {\partial \over \partial t_{2}} \; t_{2} (1 -
t_{2})  \;  {\partial \over \partial t_{2}} \;   - t_{2} \; {( - H
- ie ) \over 2}    - {1 \over 4 t_{2}} \; {\partial^{2} \over
\partial \phi^{2}} \;  ,
\label{6.2}
\end{eqnarray}

\noindent or after substituting the expression  for  $H$
\begin{eqnarray}
\hat{B} = -ie \; {t_{1} + t_{2}  - 2t_{1} t_{2} \over t_{1}
-t_{2}}  +   {2t_{2}(1-t_{1})(1 - 2 t_{1}) \over t_{1} -
t_{2}}  {\partial \over \partial t_{1}}  +
{2t_{1}(1-t_{2})(1 - 2 t_{2}) \over t_{2} - t_{1}} \;
 {\partial \over \partial t_{2}}
\nonumber
\\
+  {2 t_{1} t_{2} (1 - t_{1})^{2} \over t_{1} - t_{2}} \;
{\partial^{2} \over  \partial t_{1}^{2}}  +  {2 t_{2} t_{1} (1
- t_{2})^{2} \over t_{2} - t_{1}}   {\partial^{2} \over
\partial t_{2}^{2}}  -  {t_{1} + t_{2}  \over  2 t_{1} t_{2}}
 {\partial^{2} \over \partial \phi^{2} } \;  ;
\label{6.3}
\end{eqnarray}
\end{widetext}

\noindent note the identity

\begin{eqnarray}
-ie \; {t_{1} + t_{2}  - 2t_{1} t_{2} \over t_{1} -t_{2}}  = -ie\;
\cos \theta = -ie\; {q_{3} \over q } \; .
\nonumber
\end{eqnarray}

 Now we
turn
 to establishing connection between $\hat{B}$ and
Runge -- Lenz vector. It is convenient to solve the task  in the
same time  both in space $H_{3}$ and $S_{3}$.

In  $H_{3}$ and $S_{3}$ the quantum mechanical Runge-Lenz operator
is constructed from  momentum and orbital momentum by the formula
\cite{1979-Higgs, 1979-Leemon, 1979-Kurochkin-Otchik, 1980-Bogush-Kurochkin-Otchik}
\begin{eqnarray}
\vec{A} = e{\vec{q} \over q} + {1\over 2} \; ([\vec{L} \;\vec{P}]
- [\vec{P} \;\vec{L}]) \; ,
\label{6.7}
\end{eqnarray}

\noindent where
\begin{eqnarray}
 P_{i} =  -i (\delta_{ij}  \mp q_{i} q_{j})
\; {\partial \over
 \partial q_{j}} \; ,
\;
  \vec{L} = [ \vec{q} \;\vec{P}] \;\; ,
\label{6.8}
\end{eqnarray}

\noindent  upper sign corresponds to the model  $H_{3}$, lower
corresponds to
 $S_{3}$ model; operators  $\vec{L}$ and  $\vec{P}$  are measured in units
 $\hbar$ and  $\hbar / \rho$ respectively.

In correspondence with symmetry of  space models, the components
of
 $\vec{P}, \; \vec{L}$ obey the commutation relations of
 Lie algebras  $so(3.1)$ and $so(4)$:
\begin{eqnarray}
[L_{a}, \; L_{b}] =i\; \epsilon _{abc} \; L_{c}\; ,\;\;
 [L_{a},
\; P_{b}] =i \; \epsilon _{abc} \; P_{c}\; ,\;\; [P_{a},
P_{b}] =\pm i\; \epsilon _{abc}\; L_{c}\; .
\label{6.9}
\end{eqnarray}

\noindent Since in  the above expression for $\hat{B}$ specific
term $\;-ie\;q_{3} / q \; $  is present,  (in the model  $H_{3}$
we see the term
 $\; e\;q_{3} / q \;$), it is natural to look for certain relationship   between
 $\hat{B}$ and  $A_{3}$.

 Rather long calculation give the following result
 \begin{eqnarray}
\mbox{in} \;\;\; H_{3} \; , \qquad \qquad \hat{B} =  (\; A \; + \;
\vec{L}^{2}\; ) \; ,\qquad
\mbox{in} \;\;\; S_{3} \; , \qquad       \qquad i \; \hat{B} = (\;
A \; + \;i\; \vec{L}^{2}\; ) \; .
\label{6.33}
\end{eqnarray}

\section{Discussion}


In should be emphasized that the possibility to employ complex-valued coordinates in space of positive
 constant curvature $S_{3}$ can be used in other coordinate systems as well -- Olevsky's results
 provide us with 34-6 = 28  such special cases. For instance,
 a complex analogue in spherical space $S_{3}$  for horospherical coordinates of
 Lobachevsky space was introduced in \cite{Ovsiyuk, Bychkovskaya}, ant it was used to
  examine Shapiro's plane wave
  solutions of the Schr\"{o}dinger equation
   in spaces $S_{3}$ by analogy with  $H_{3}$ model.

Such a possibility evidently  will extend the class  of integrable problems in
these spaces -- see \cite{Herranz-Ballesteros-2006}.
 Also, complex coordinates  in 3D-spaces of constant curvature
can be of interest  in the context of the theory of the Lorentz  group $SO(3,1)$ -- see in
\cite{Bogush-Red'kov, Red'kov-Bogush-Tokarevskaya}.

\section{Acknowledgement}

Authors are grateful to participants of the  scientific  seminar of Laboratory of theoretical physics of Institute
of physics, National academy of sciences of Belarus,  for discussion and advices.

\label{last}

\end{document}